\documentclass[ 
                superscriptaddress,
                twocolumn,
                longbibliography
                ]{revtex4-2}

\usepackage{hyperref} 
\usepackage{amsmath}
\usepackage{amssymb}
\usepackage{graphicx}
\usepackage{listings} 
\usepackage{placeins} 
\usepackage{xcolor}
\usepackage[normalem]{ulem}

\usepackage{listings}
\usepackage{xcolor}

\definecolor{codegreen}{rgb}{0,0.6,0}
\definecolor{codepurple}{rgb}{0.58,0,0.82}
\definecolor{backcolour}{rgb}{0.95,0.95,0.92}

\lstdefinestyle{mystyle}{
  backgroundcolor=\color{backcolour},   
  commentstyle=\color{codegreen},
  keywordstyle=\color{orange},
  stringstyle=\color{codepurple},
  basicstyle=\ttfamily\footnotesize
}

\usepackage[utf8]{inputenc}
\usepackage[T1]{fontenc}

\newcommand{\ket}[1]{| #1 \rangle}
\newcommand{\bra}[1]{\langle #1 |}
\newcommand{\braket}[2]{\langle #1 | #2 \rangle}

\newcommand{\identity}{1\!\!1}

\newcommand{\innsbruck}{Institute for Theoretical Physics, University of Innsbruck, A-6020 Innsbruck, Austria}
\newcommand{\parityqc}{Parity Quantum Computing GmbH, A-6020 Innsbruck, Austria}
\newcommand{\rle}{Research Laboratory of Electronics, Massachusetts Institute of Technology, Cambridge, MA 02139, USA}
\newcommand{\mitphysics}{Department of Physics, Massachusetts Institute of Technology, Cambridge, MA 02139, USA}
\newcommand{\harvardphysics}{Department of Physics, Harvard University, Cambridge, MA 02138, USA}
\newcommand{\miteecs}{Department of Electrical Engineering and Computer Science, Massachusetts Institute of Technology, Cambridge, MA 02139, USA}
\newcommand{\mitll}{MIT Lincoln Laboratory, 244 Wood Street, Lexington, MA 02420, USA}

\begin{document}

\title[]{CircuitQ: An open-source toolbox for 
superconducting circuits}

\author{Philipp Aumann}
\email{philipp.aumann@uibk.ac.at}
\affiliation{\innsbruck}

\author{Tim Menke}
\affiliation{\rle}
\affiliation{\mitphysics}
\affiliation{\harvardphysics}

\author{William D. Oliver}
\affiliation{\rle}
\affiliation{\mitphysics}
\affiliation{\mitll}
\affiliation{\miteecs}

\author{Wolfgang Lechner}
\email{wolfgang.lechner@uibk.ac.at}
\affiliation{\innsbruck}
\affiliation{\parityqc}

\date{\today}

\begin{abstract}
We introduce CircuitQ, an open-source toolbox for the analysis of superconducting circuits implemented in Python.
It features the automated construction of a symbolic Hamiltonian of the input circuit and a dynamic numerical representation of the Hamiltonian with a variable basis choice.
The software implementation is capable of choosing the basis in a fully automated fashion based on the potential energy landscape.
Additional features include the estimation of the $T_1$ lifetimes of the circuit states under various noise mechanisms.
We review previously established circuit quantization methods and formulate them in a way that facilitates the software implementation.
The toolbox is then showcased by applying it to practically relevant qubit circuits and comparing it to specialized circuit solvers.
Our circuit quantization is applicable to circuit inputs from a large design space, and the software is open-sourced.
We thereby add an important resource for the design of new quantum circuits for quantum information processing applications.
\end{abstract}

\maketitle

\section{Introduction}

\begin{figure*}[t]
\includegraphics{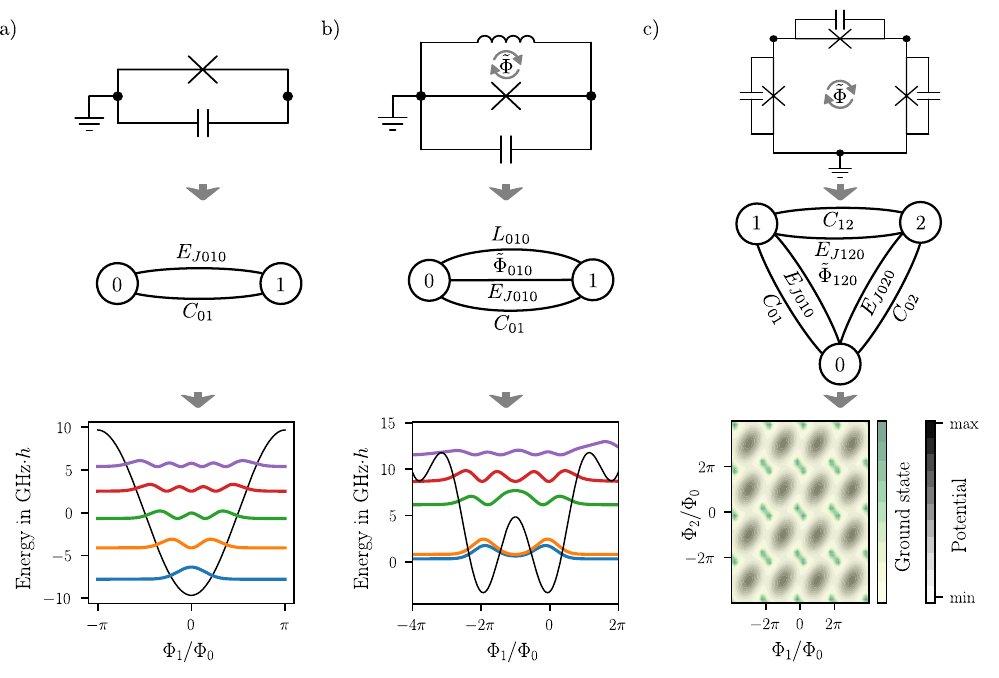}
\caption{Overview of the functionality of CircuitQ for three example circuits, as discussed in section~\ref{sec:demonstration}.
The general procedure to analyze a superconducting circuit (top row) is to interpret it as a graph whose edges correspond to the circuit elements (middle row).
The graph serves as the input of the toolbox, and a symbolic and a numerical Hamiltonian are constructed from it.
The potential and eigenspectrum of the Hamiltonian are shown in the bottom row.
The symbolic Hamiltonians and parameter values can be found in section~\ref{sec:demonstration}, and the code to generate these instances is listed in Appendix~\ref{appendix:code_fig_examples}. 
(a) Transmon circuit. The last row depicts the scaled absolute value squared of the wavefunctions of the lowest five eigenstates as a function of the flux variable $\Phi_1$, which is the node flux variable connected to node $1$ of the respective graph in the middle row, while the scaling factor $\Phi_0$ is the flux quantum.
Each wave function is offset by its corresponding eigenvalues.
The potential of the Hamiltonian is shown as a black line.
(b) Fluxonium circuit. The scaled wavefunctions of the eigenstates and potential energy are shown in the same fashion as in (a).
(c) Persistent-current flux qubit circuit.
Due to the additional node, the problem becomes two-dimensional. Similar to before, $\Phi_1$ and $\Phi_2$ are the node variables of node $1$ and $2$ in the circuit graph.
We show an overlap of two contour plots, where one depicts the scaled absolute value squared of the wavefunction of the lowest eigenstate and the other visualizes the potential.}
\label{fig:examples}
\end{figure*}

Superconducting circuits are one of the most versatile and promising platforms in the development of chip-based quantum processors \cite{arute_quantum_2019, blais2020circuit}.
The development of cutting-edge qubit designs is currently driven by the effort to realize quantum computers with long coherence times as well as high-fidelity control and readout, all in a scalable design \cite{kjaergaard_superconducting_2020}.
Combining these requirements is a grand challenge for quantum hardware design.
Therefore, considerable effort is invested into the study of new and improved qubit circuits for quantum information processing applications~\cite{weiss_spectrum_2019, paolo_control_2019, nguyen_high-coherence_2019, mirrahimi_dynamically_2014}.

A major part of the analysis of superconducting circuits is the construction of a quantum model to describe the system theoretically.
Such a model is obtained by using general methods to construct the corresponding Hamiltonian~\cite{burkard_multilevel_2004, vool_introduction_2017, kerman_efficient_2020}.
A numerical implementation of the algebraic description is then needed to analyze the quantum properties of the circuit.
A number of open source software packages has been developed for this purpose.
The library scQubits \cite{groszkowski_scqubits_2021}, for example, simulates qubits from a specific set of circuits.
The package QuCAT \cite{gely_qucat_2020} offers a more general circuit input, as it permits a combination of Josephson junctions, inductances, capacitances and resonators. The quantization is performed in the basis of normal modes, which is suitable for weakly anharmonic systems.
Another useful toolbox is provided by the SuperQuantPackage~\cite{andrey_klots_andreyklotssuperquantpackage_2021}, which includes an algorithm to provide the user with a numerical representation of the Hamiltonian for a given input circuit. It performs a coordinate transformation prior to quantization.
Qiskit Metal~\cite{qiskit_2022} and KQCircuits~\cite{noauthor_iqm-finlandkqcircuits_2022} enable the analysis of superconducting circuits based on their physical layout on the chip.
While such software packages have been proven to be useful for specific circuit design tasks, we expand on prior work by presenting a toolbox that works for a generic variety of circuits, determines a symbolic and numerical Hamiltonian, provides an automated choice of implementation basis and includes a measure for several $T_1$ contributions.

In this work, we provide a structured review of the superconducting circuit quantization procedure implemented in the software toolbox CircuitQ.
This provides an insight into the software implementation, but also serves as a more general review of the quantization process of superconducting circuits by constructing the Hamiltonian.
CircuitQ is written in Python and can be used to analyze superconducting circuits that are a user-defined combination of Josephson junctions, linear inductances, and capacitances.
It takes the circuit and optionally the circuit component parameters as an input and returns the quantum physical properties of the circuit, particularly the corresponding Hamiltonian in symbolic and numerical form. Figure~\ref{fig:examples} provides an overview of this conceptual procedure for three example circuits.
Depending on the shape of the inductive potential, CircuitQ can dynamically perform the numerical implementation in the charge basis, the flux basis, or in a mixture of both. Therefore, it provides a toolbox for circuits comprising different parameter regimes. As detailed in section~\ref{sec:demonstration}, this implementation works well for few-node circuits, where the direct implementation of the node variables with the flux and charge basis is a natural choice, whereas the limits of this implementations are reached for more complex circuits.
The interface offers the possibility for a general circuit input.
A variety of features and degrees of freedom can be adjusted by the user, such as the circuit composition, the component parameters, ground nodes, offset charges, and loop fluxes.
In order to analyze the circuit in view of noisy environments, we implemented an estimation of the $T_1$ lifetime of an eigenstate by considering three common relaxation mechanisms.
Finally, we showcase the software by comparing its accuracy to specialized circuit solvers for several prominent qubit circuits.
We provide community access to CircuitQ by making the code and documentation publicly available on GitHub~\footnote{Link to GitHub repository: \url{https://github.com/PhilippAumann/circuitq}}.

The article is organized as follows.
In section~\ref{sec:symbolic_Hamiltonian}, we present the procedure to generate the symbolic Hamiltonian from a given input circuit.
Subsequently, the numerical implementation of the Hamiltonian and features of the toolbox are presented in section~\ref{sec:numerical_Hamiltonian}.
Lastly, applications of the toolbox to practically relevant circuit examples are provided in section~\ref{sec:demonstration}.

\section{From the circuit to the symbolic Hamiltonian}\label{sec:symbolic_Hamiltonian}

\begin{figure}[t!]
\includegraphics{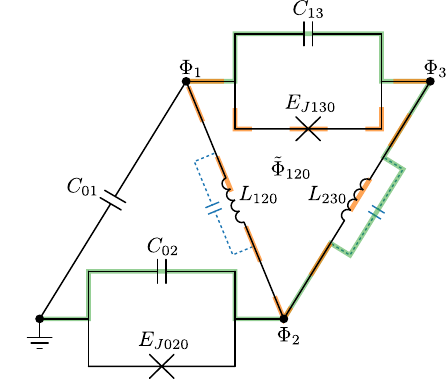}
\caption{Example circuit to demonstrate the procedure of constructing a Hamiltonian from a superconducting circuit graph. Parasitic capacitances are indicated by the blue elements. A possible spanning tree is indicated in green and a closed inductive loop is indicated by the orange dashed lines. We provide a detailed overview of the circuit quantization steps in the main text.}
\label{fig:circuit}
\end{figure}

The first step in the analysis of the quantum properties of a superconducting circuit is the construction of the circuit Hamiltonian.
Our software implementation automates the process to derive the symbolic Hamiltonian.
A multitude of techniques for circuit quantization has been developed, including several based on the method of nodes \cite{burkard_multilevel_2004, girvin_circuit_2014, vool_introduction_2017, kerman_efficient_2020}, black box quantization~\cite{nigg2012black, solgun2014blackbox} and others \cite{minev_energy-participation_2021, minev2021circuit}.
Here we follow the method of nodes-based approach, which is generally applicable to any superconducting circuit that includes capacitances, inductances, and Josephson junctions, given the realistic condition that spurious capacitances exist between all circuit nodes.

The starting point of circuit quantization is a circuit diagram such as the one shown in figure~\ref{fig:circuit}, which is a lumped-element representation of a prospective on-chip microfabricated device.
A similar version of this circuit example can be found in reference~\cite{leib_many-body_2015}.
The circuit diagram can be seen as a graph with the circuit elements on its edges.
The nodes are then home to the conjugate charge and flux variables, which represent the charge stored on capacitances connected to the node, and the flux along a specific path from the node to ground, respectively.
They are denoted as the node charge $q_i$ and node flux $\Phi_i$, while $\vec{q}$ and $\vec{\Phi}$ are the vectors of all charge and flux variables ordered by node index.
One or more ground nodes can be specified by the user when initializing an instance of the \texttt{CircuitQ} class.
Additionally, all active nodes with only one neighbouring node are added to the ground nodes if the neighbouring node is ungrounded. An active node is a node that is connected to a capacitance as well as an inductive element.
If no ground node could be specified, an active node is chosen to be the ground node. Node $0$ is the ground node of the circuit depicted in figure~\ref{fig:circuit}.
Ground nodes do not appear in the constructed Hamiltonian and the associated variables are removed from vector $\vec{q}$ and $\vec{\Phi}$.

In the software, the circuit graph is implemented as a \texttt{MultiGraph} instance of the NetworkX package~\cite{hagberg_exploring_2008}.
A code example to initialize the circuit in figure~\ref{fig:circuit} is given in Appendix~\ref{appendix:code_fig_circuit}.
We note that a graph is automatically simplified using the common rules for parallel and series capacitors.

The definition of the node fluxes requires the choice of a unique path from each node in the circuit graph to ground.
The set of such paths for all nodes is termed the \textit{spanning tree} of the graph, which cannot contain loops.
There may be multiple ways to choose it for a given circuit.
One such choice is highlighted as a green sub-graph in figure~\ref{fig:circuit}.
The choice of spanning tree is equivalent to setting a gauge and therefore does not change the physics of a circuit \cite{vool_introduction_2017}.
However, it does affect the circuit quantization procedure and the form of the Hamiltonian.
In this work, we follow common practice and route the spanning tree through capacitive circuit elements only.
Since each node pair is connected by spurious capacitances, a spanning tree can always be defined in such a way. For the implementation, CircuitQ makes use of the spanning tree functionality of the NetworkX package.

In order to determine the contribution of an inductive circuit element to the Hamiltonian, one needs to evaluate the flux difference across the respective edge.
An inductive loop consisting of inductances and junctions can encircle an external flux, and the boundary condition has to be fulfilled that all the fluxes around a loop sum to a multiple of the flux quantum $\Phi_0 = \frac{\hbar}{2e}$.
We highlight one such loop in orange in figure~\ref{fig:circuit}.
The external flux enters the circuit Hamiltonian by being added to the flux difference across an inductive element between two of the nodes that are part of the loop.

In CircuitQ, the automated evaluation of flux differences is started by defining the set of inductive edges $\mathcal{L}$ and splitting it into the subset $\mathcal{S}$ that is in parallel to spanning tree edges and into the remaining edges $\mathcal{B} = \mathcal{L} - \mathcal{S}$.
As we allow for multiple parallel inductive elements between two nodes, a restriction needs to be introduced that only the first parallel inductive edge is added to $\mathcal{S}$.
The routine then iteratively steps through the edges in $\mathcal{B}$.
If the current edge does not close an inductive loop, we add it to a subset $\mathcal{B}_\text{o} \subseteq \mathcal{B}$ and do not assign an external flux to it.
In case it does close an inductive loop, we add it to $\mathcal{B}_\text{c} \subseteq \mathcal{B}$ and do assign an external flux to it.
Using the fluxoid quantization condition, the directionality of the edges in $\mathcal{S}$ can be chosen such that we obtain the following relation for the edge flux $\Phi_{e_{ijn}}$ of the $n$-th inductive edge $e_{ijn}$ between nodes $i$ and $j$:
\begin{align}\label{eq:edge_flux}
    \Phi_{e_{ijn}} =
    \begin{cases}
        \Phi_j - \Phi_i & \textrm{for } e_{ijn} \in \mathcal{S} \cup \mathcal{B}_\text{o} \\
        \Phi_j - \Phi_i + \tilde{\Phi}_{ijn}  & \textrm{for } e_{ijn} \in \mathcal{B}_\text{c} \\
    \end{cases}.
\end{align}
For the circuit in figure~\ref{fig:circuit}, this procedure identifies the loop flux $\tilde{\Phi}_{120}$ and applies it to the flux difference between nodes 1 and 2.

We note that the loop fluxes are degrees of freedom that can be used to tune the circuit properties.
At the same time, they open a path for undesired fluctuations from the environment to couple to the circuit. CircuitQ determines a set of loop fluxes automatically when initializing an instance for a given circuit. These fluxes are treated as conventional circuit parameters whose numerical values can be specified by the user.

Given the relationship between branch and node fluxes, we can explicitly state the inductive potential of the Hamiltonian, which is the sum of linear inductive and Josephson potentials:
\begin{align}\label{eq:potential_energy}
    U_\text{ind}\left(\vec{\Phi}\right) = &\sum_{ \{i, j\} \,\in\, \mathcal{I}_\text{L}} \sum_{n=0}^{N_\text{L}^{ij}-1} \frac{\left(\Phi_{e_{ijn}}\left(\vec{\Phi}\right)\right)^2}{2L_{ijn}} \nonumber \\
    & -\sum_{\{i, j\} \,\in\, \mathcal{I}_\text{J}} \sum_{n=0}^{N_\text{J}^{ij}-1} E_{Jijn} \cos\left( \frac{\Phi_{e_{ijn}}\left(\vec{\Phi}\right)}{\Phi_0} \right), 
\end{align}
where $N^{ij}_L$ is the number of linear inductors between node $i$ and $j$ and $N^{ij}_J$ is the number of Josephson junctions, while $\mathcal{I}_\text{L}$ and $\mathcal{I}_\text{J}$ represent the set of node pairs connected by inductors and Josephson junctions.
Our formulation of the potential expands upon prior work on general circuit quantization formulations in that it allows for multiple parallel inductive elements per node pair.
While this is relevant for Josephson junctions, for example in modelling a frequency tunable transmon, where two Josephson junctions form a SQUID loop, multiple parallel linear inductors are of limited practical relevance for quantum information processing applications because they lead to sensitivity of the circuit energy to static flux offsets.

The kinetic energy $T_\text{cap}$ of the Hamiltonian is given by the capacitive energy of the circuit.
In general, $T_\text{cap}$ takes the form
\begin{equation}
    T_\text{cap} \left( \vec{q} \right) = \frac{1}{2} \vec{q}^{\text{T}}  \textbf{C}^{-1} \vec{q}.
\end{equation}
Here, $\textbf{C}$ is the node capacitance matrix, which contains the sum of all capacitances connected to a node as the respective diagonal entry and the negative capacitance between two nodes on the off-diagonals. 
That is, the $i$-th diagonal element of the matrix is given by
\begin{align}
    \textbf{C}_{i,i} &= \sum_n C_{in}
\end{align}
by summing over all capacitances $C_{in}$ connected to node $i$, while the off-diagonal elements are given by
\begin{align}
    \textbf{C}_{i,j} &= - C_{ij}
\end{align}
for $i \neq j$ and capacitance $C_{ij}$ linking nodes $i$ and $j$. The capacitances are invariant under permutation of the indices: $C_{ij} = C_{ji}$.
We note that the rows and columns of $\textbf{C}$ that are associated to ground nodes are removed from the matrix.
The form of the kinetic energy arises in the Legendre transformation of the circuit Lagrangian.
Charge offsets on a node are taken into account by directly adding the offset to the respective charge operator.

The circuit Hamiltonian in terms of the conjugate coordinates $\vec{\Phi}$ and $\vec{q}$ is given by the sum of the kinetic and potential terms:
\begin{equation}\label{eq:symbolic_hamiltonian}
    H\left(\vec{\Phi}, \vec{q} \right) = T_\text{cap}\left( \vec{q} \right) + U_\text{ind}\left(\vec{\Phi}\right).
\end{equation}
In CircuitQ, it is returned as a symbolic SymPy object~\cite{meurer_sympy_2017}. The Hamiltonian that is constructed for the example circuit in figure~\ref{fig:circuit} can be found in Appendix~\ref{appendix:leib_circuit_hamiltonian}.

For the quantum physical treatment of the system described by the Hamiltonian in equation~\ref{eq:symbolic_hamiltonian}, we perform the usual quantization procedure by promoting the conjugate variables to operators:
\begin{align}
    \Phi_i \rightarrow \hat{\Phi}_i \ \forall \Phi_i \in \vec{\Phi} \ \textrm{ and } q_i \rightarrow \hat{q}_i\ \forall q_i \in \vec{q} .
\end{align}
Those operators fulfill the canonical commutation relations
\begin{align}
    [\hat{\Phi}_i, \hat{q}_j] = i \hbar \delta_{i,j}.
\end{align}
Generating the Hamiltonian in a symbolic form is the first step towards the automated description of a circuit. The second step is to express the quantized Hamiltonian numerically.

\section{From the symbolic Hamiltonian to its numerical implementation}
\label{sec:numerical_Hamiltonian}

The numerical implementation of the symbolic Hamiltonian is crucial for the analysis of the quantum properties of the input circuit. In this section, we first present important steps of the automated numerical implementation that is part of the toolbox. Subsequently, we describe numerical analysis tools that are implemented in CircuitQ.

\subsection{Implementation}

Numerical values for the circuit parameters can be specified individually.
This includes capacitances, inductances, Josephson energies, external fluxes and charge offsets. If the values are not specified by the user, they are set to a default value.
In addition to the numerical parameter values, the charge and flux operators that appear in the Hamiltonian have to be implemented as numerical matrices.
These matrices can be either formulated in the flux basis or in the charge basis, which is analogous to choosing a position or momentum space representation.
If the potential is periodic along the direction of a node flux variable, the charge basis is the preferred choice for implementing the variables of this node.
The connection of a linear inductance to a node leads to a non-periodic harmonic contribution of the potential, which makes a flux basis implementation for the corresponding node variables more desirable.
To distinguish these two cases, we label a node as periodic if there is no linear inductance connected to it and if neighbouring nodes that are connected via a Josephson junction are periodic as well.
Periodic node variables are automatically implemented in the charge basis and non-periodic variables in the flux basis.

\subsubsection{Flux Basis}
To implement a node variable in the flux basis, we confine the numerical flux values to a finite grid ${[-\Phi_{\text{max}}, -\Phi_{\text{max}} + \delta, \dots, \Phi_{\text{max}}]}$ with grid spacing $\delta$. The grid length can be decided by the user or is set to a default value otherwise. Consequently, we can assign a diagonal matrix to the flux variable:
\begin{align}
    \hat{\Phi} \rightarrow    
    \left( 
        \begin{matrix}
            -\Phi_{\text{max}} & & & \\
            & -\Phi_{\text{max}} + \delta & & \\
            & & \ddots & \\
            & & & \Phi_{\text{max}}
        \end{matrix}
    \right).
\end{align}
As the conjugate momentum of the flux, the charge variable can be associated with the derivative with respect to $\Phi$: ${q = -i \hbar \partial_{\Phi}}$. To implement the derivative as a Hermitian operator, we use the finite difference method:
\begin{align}
    \hat{q} \rightarrow  \frac{-i \hbar}{2 \delta}  
    \left( 
        \begin{matrix}
            0 & 1 & & \\
            -1& 0 & 1 & \\
            & & \ddots & \\
            & & -1 & 0
        \end{matrix}
    \right),
\end{align}
\begin{align}
    \hat{q}^2 \rightarrow  \frac{- \hbar^2}{\delta^2}  
    \left( 
        \begin{matrix}
            -2 & 1 & & \\
            1& -2 & 1 & \\
            & & \ddots & \\
            & & 1 & -2
        \end{matrix}
    \right).
\end{align}
To generate Hermitian matrices, we have chosen a different discretization for the first and second derivative.
The cosine terms in equation~\ref{eq:potential_energy}, referring to the energy contribution of Josephson junctions, can be represented by diagonal matrices in the flux basis, where the diagonal elements are the cosine of the corresponding numerical edge flux value.
CircuitQ is capable of working with charge and flux offsets, where the flux offsets $\tilde{\Phi}$ are associated with loop fluxes and charge offsets $\tilde{q}$ with node charges. They are implemented by multiplying them with the identity matrix: 
\begin{align}
    \tilde{q} \rightarrow  \tilde{q} \cdot \identity, \ \ \tilde{\Phi} \rightarrow \tilde{\Phi} \cdot \identity.
\end{align}
A Hilbert space is assigned to every node which is not set to ground.
To obtain a numerical description of the full Hamiltonian, these subspaces are combined into a composite space using the tensor product by substituting
\begin{align}
    \hat{\Phi}_i &\rightarrow \identity \otimes \cdots \otimes \identity \otimes \underset{\substack{\downarrow \\ \text{position of node } i}}{\hat{\Phi}} \otimes \identity \otimes \cdots, \\
    \hat{q}_i &\rightarrow \identity \otimes \cdots \otimes \identity \otimes \underset{\substack{\downarrow \\ \text{position of node } i}}{\hat{q}} \otimes \identity \otimes \cdots .
\end{align}
Here, the variable corresponding to node $i$ is implemented by placing the respective matrix at the position of the composite space which corresponds to node $i$.
We note that node numbering may change due to the elimination of the ground nodes.
The sequence of the nodes is deduced by the algorithm and kept consistent throughout the evaluation of an instance. 

\subsubsection{Charge Basis}\label{subsubsec:charge_basis}
Similar to the flux basis, we restrict the charge variables to a finite grid when using the charge basis. The charge is truncated at a cutoff number of Cooper pairs $n_{\text{cutoff}}$, which leads to the charge grid ${2e[-n_{\text{cutoff}}, \dots, n_{\text{cutoff}}]}~=~{[-q_{\text{\text{max}}}, \dots, q_{\text{\text{max}}}]}$. In this setting, we can express the charge variable as a diagonal matrix:
\begin{align}
    \hat{q} \rightarrow  2e  
    \left( 
        \begin{matrix}
            -n_{\text{cutoff}} & & & \\
            & -n_{\text{cutoff}} + 1 & & \\
            & & \ddots & \\
            & & & n_{\text{cutoff}}
        \end{matrix}
    \right).
\end{align}
Flux variables that correspond to periodic nodes appear exclusively in the arguments of the cosine terms in the Hamiltonian. The cosine acts as a hopping operator in the Cooper pair number basis~\cite{langford_circuit_2013}:
\begin{align}\label{eq:cos_in_cooper_pair_basis}
    \cos \left( \frac{\hat{\Phi}}{\Phi_0}\right) \rightarrow \frac{1}{2} \sum_n \ket{n} \bra{n+1} + \ket{n+1} \bra{n},
\end{align}
with $\ket{n}$ being the charge state corresponding to $n$ Cooper pairs.
We can then use the decomposition of the cosine into complex exponentials, 
\begin{align}
\cos \left( \frac{\hat{\Phi}}{\Phi_0} \right) = \frac{1}{2} \left( e^{i\frac{\hat{\Phi}}{\Phi_0}}+e^{-i\frac{\hat{\Phi}}{\Phi_0}}\right),
\end{align}
to represent these terms numerically. This procedure is used in scQubits~\cite{groszkowski_scqubits_2021}.
The exponential of a flux operator describes the tunneling process of a Cooper pair through a Josephson junction, and it can be described as a jump operator in the charge basis~\cite{langford_circuit_2013}:
\begin{align}
    e^{i\frac{\hat{\Phi}}{\Phi_0}} \rightarrow   
    \sum_n \ket{n} \bra{n+1}
    \rightarrow
    \left( 
        \begin{matrix}
            0 & & & \\
            1 & 0 & & \\
            & & \ddots & \\
            & & 1& 0
        \end{matrix}
    \right).
\end{align}
The composite space has to be considered if there are multiple flux variables in the argument of the cosine. In this case, we again make use of the tensor product:
\begin{align}\label{eq:exponential_implementation}
    e^{i\frac{ \hat{\Phi}_i - \hat{\Phi}_j}{\Phi_0} } \rightarrow \identity \otimes \cdots \otimes \hspace{-.5cm} \underset{\substack{\downarrow \\ \text{position of node } i}}{e^{i\frac{\hat{\Phi}}{\Phi_0}}} \hspace{-.5cm} \otimes  \cdots \otimes \hspace{-.5cm} \underset{\substack{\downarrow \\ \text{position of node } j}}{\left(e^{i\frac{\hat{\Phi}}{\Phi_0}}\right)^\dagger} \hspace{-.5cm} \otimes \identity \otimes \cdots .
\end{align}
The full cosine function can thus be implemented as
\begin{align}\label{eq:cos_implementation}
    \cos\left(\frac{ \hat{\Phi}_i - \hat{\Phi}_j}{\Phi_0} \right) \rightarrow \frac{1}{2} \left( e^{i\frac{\hat{\Phi}_i - \hat{\Phi}_j}{\Phi_0}} + \left( e^{i\frac{\hat{\Phi}_i - \hat{\Phi}_j}{\Phi_0}}\right)^{\dagger} \right).
\end{align}
To account for a flux offset $\tilde{\Phi}$, the exponential function in equation~\ref{eq:exponential_implementation} can be multiplied by the complex scalar $e^{-i\tilde{\Phi}}$.

The choice of an appropriate numerical value for the discretization parameters $\Phi_{\text{max}}$, $\delta$ and $n_{\text{cutoff}}$, which determine the numerical representation of the flux and charge variables, depend on the particular circuit. An appropriate regime can be found by increasing (for the case of $\Phi_{\text{max}}$ and $n_{\text{cutoff}}$) or decreasing (for the case of $\delta$) the numerical value of the parameter until convergence is reached, such that the resulting spectrum of the circuit becomes almost invariant under a slight modification of those values.  

The numerical grid for the numerical Hamiltonian is generated using the \texttt{lambdify} function of SymPy~\cite{meurer_sympy_2017} with the parameters and matrices that have been described in this section as inputs.
The final implementation is returned as a sparse matrix in SciPy format~\cite{virtanen_scipy_2020}.

For some analyses, it may be helpful to visualize the eigenstates as a function of the flux variable even when an implementation in the charge basis has been used.
For this purpose, CircuitQ provides a method which transforms the eigenvectors from the charge to the flux basis. To transform a state vector, given in the charge basis $\mathcal{B}_q = \{\ket{q_i} \}_i$, to a representation in the flux basis $\mathcal{B}_{\Phi} = \{\ket{\Phi_i}\}_i$, the transformation matrix $T$ can be defined, which maps the state vector from the charge to the flux basis. The coefficients of this matrix read:
\begin{align}
    T_{i,j} = \braket{q_i}{\Phi_j} = \dfrac{1}{\sqrt{d}} e^{-\frac{i}{\hbar} q_i \Phi_j},
\end{align}
with $d$ being the number of basis vectors.
We follow the same procedure in our numerical implementation, however we use a modified transformation matrix which respects the construction of the composite Hilbert space, which, in general, consists of subspaces that are either implemented in the charge or the flux basis. We note that depending on the size of the numerical matrices, this transformation can be numerically demanding and consequently may lead to a bottleneck in computation time.

\subsection{Features for circuit analysis}\label{subsec:features}
Since CircuitQ constructs a numerical implementation of the circuit Hamiltonian, it can be used as a tool for the analysis of the quantum properties of superconducting circuits. This includes the energy spectrum of the Hamiltonian and relaxation times of the energy eigenstates.

\begin{table*}[thb]
\centering
\linespread{1.5}\selectfont
\caption{\linespread{1.1}\selectfont Overview of the noise contributions to $T_1$ which are implemented in CircuitQ. The second column lists the formulas used as the basis of the noise estimation. Here, $\ket{g}$ and $\ket{e}$ label the qubit ground and excited state. The angular frequency of the qubit is labeled by $\omega_q$. Other symbols are described in the third column.}
\label{tab:T_1}
\begin{tabular}{p{2cm} p{7.5cm} p{8cm} }
	\hline 
Contribution	& Formula & \\ 
\hline \hline
Quasiparticle tunneling~\cite{catelani_relaxation_2011} & \vspace{0.4cm} \begin{equation}
\begin{aligned} \label{eq:T1_qp} 
T_{1_{\textrm{qp}}} = \Biggl( \frac{S_{\textrm{qp}}(\omega_q)}{\hbar^2} \biggl( & \sum_j E_{J, j} \left| \langle g| \sin \left( \frac{\hat{\Phi}_{e_{j}}}{2\Phi_0}\right) | e \rangle \right|^{2} \\ 
& + \sum_l E_{L, l} \left| \langle g| \frac{\hat{\Phi}_{e_{l}}}{2\Phi_0} | e \rangle \right|^{2} \bigg) \Bigg)^{-1}\end{aligned}\end{equation} & 
$S_{\textrm{qp}} (\omega_q) = \hbar x_{\textrm{qp}}\frac{8}{\pi}\sqrt{\frac{2 \Delta}{\hbar \omega_q}}$: Noise spectral density~\cite{catelani_relaxation_2011, nguyen_high-coherence_2019} \newline 
$\hat{\Phi}_{e_{j/l}}$: Edge flux operator describing the flux of the inductive edge corresponding to the $j$-th Josephson junction or $l$-th linear unductance (see equation~\ref{eq:edge_flux}) \newline
$E_{J,j}:$ Josephson energy of the $j$-th junction \newline
$E_{L,l} = \frac{\Phi_0^2}{L_l}:$ Inductive energy of the $l$-th linear inductance \newline
$x_{\textrm{qp}}= 10^{-8}$: Density of quasiparticles which is scaled by the density of Cooper-pairs~\cite{nguyen_high-coherence_2019} \newline
$\Delta = 1.76 \cdot k_B T_c$: Superconducting gap~\cite{fernandes_lecture_nodate} \newline
$T_c = 1.2$ K: Critical temperature of aluminum~\cite{cochran_superconducting_1958} \\
\hline 
Dielectric loss~\cite{nguyen_high-coherence_2019} & \vspace{0.3cm}\begin{equation}\label{eq:T1_diel} T_{1_\textrm{diel}} = \left( \sum_{i} \frac{S_{Q_i}(\omega_q)}{\hbar^2} \left| \bra{g} \hat{q}_{e_i} \ket{e} \right|^{2} \right) ^{-1}\end{equation}
& $S_{Q_i}(\omega_q)= \frac{\hbar}{Q_{\textrm{cap}}(\omega_q) C_i} \left(1 + \coth \frac{\hbar \omega_q}{2k_B T} \right)$: Noise spectral density~\cite{nguyen_high-coherence_2019, smith_superconducting_2020} \newline
$\hat{q}_{e_i} = \hat{q}_{i_2} - \hat{q}_{i_1}$: Charge operator of the capacitive branch $e_i$ linking nodes $i_2$ and $i_1$ with corresponding capacitance~$C_i$ \newline
$Q_{\textrm{cap}}(\omega_q) = 3\cdot 10^6 \left( \frac{2 \pi \cdot 6 \textrm{ GHz}}{\omega_q}\right)^{0.7}$: Dielectric quality factor~\cite{smith_superconducting_2020, pop_coherent_2014} \newline
$T=15$ mK: Assumed temperature of a sample
\\
\hline
Flux noise~\cite{yan_flux_2016, nguyen_high-coherence_2019} & \vspace{0.3cm}\begin{equation}\label{eq:T1_flux}
T_{1_{\textrm{flux}}} =  \left( \sum_{i} \frac{S_{\Phi}(\omega_q)}{\hbar^2} \left| \bra{g} \hat{I}_i \ket{e} \right|^{2} \right)^{-1}\end{equation}
& $S_{\Phi}(\omega_q)= \frac{\hbar^2}{(2e)^2 \Phi_0^2} 2\pi \frac{A^2}{\omega} $: Noise spectral density~\cite{nguyen_high-coherence_2019}  \newline
$A=2\pi 10^{-6} \Phi_0$: Noise amplitude~\cite{nguyen_high-coherence_2019} \newline
$\hat{I}_i$: Current operator, which includes all $\frac{\Phi_k-\Phi_l}{L_{kl}}$ and $I_{C, kl}  \sin \frac{\Phi_k-\Phi_l}{\Phi_0}$ terms in the circuit that correspond to a circuit graph edge $i \in \mathcal{B}_c$ connecting nodes $k$ and $l$ \newline
$I_{C, kl} = \frac{2e}{\hbar}  E_{J, kl}$: Critical current \newline 
$E_{J, kl}$: Josephson energy linking node $k$ and $l$\\
\hline
\end{tabular} 
\end{table*}

\subsubsection{Spectrum}

To calculate the energy spectrum of the numerical Hamiltonian, the toolbox provides a method that returns the lowest eigenstates and eigenenergies of the numerical Hamiltonian matrix.
We use the SciPy library for the (partial) diagonalization, which in turn makes use of efficient ARPACK routines~\cite{lehoucq_arpack_1998}. With this functionality, the toolbox can be used to investigate how a change of parameter values -- for example an external flux -- or a change in the circuit composition affects the energy spectrum and eigenstates. 

For the description of the superconducting circuit as a qubit, we associate the lowest two energy levels that have a nonvanishing energy difference with the qubit states $\ket{0}$ and $\ket{1}$ by default. However, it is possible to declare a different state as the excited qubit state manually. The corresponding energy levels should not be degenerate. To operate a circuit as a qubit, a high degree of anharmonicity of its spectrum is desired. CircuitQ provides a method which gives an estimate of the harmonicity of the spectrum in a quantified form.

\begin{figure}[t]
\includegraphics{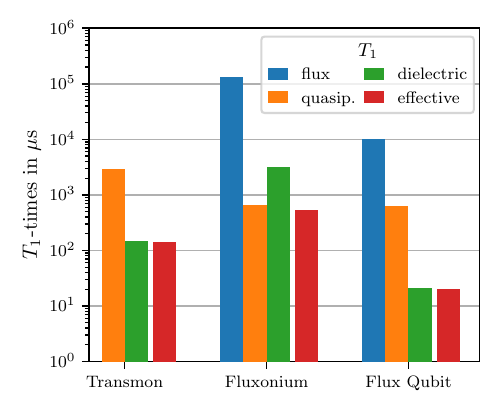}
\caption{Summary of the estimated $T_1$ times for the three circuits shown in figure~\ref{fig:examples}. The effective $T_1$ time is the result of the noise contributions due to quasiparticle tunnling, dielectric loss and flux noise (see table~\ref{tab:T_1}). The chosen parameter values for calculating the estimations are given in table~\ref{tab:values_T1} in Appendix~\ref{appendix:values_T1}. We note that these values differ from those chosen for figure~\ref{fig:examples}. There is no flux noise contribution to the transmon as we consider the fixed-frequency transmon circuit here.}
\label{fig:t1}
\end{figure}
\subsubsection{Relaxation time $T_1$}

In order to determine the performance of the circuit as a qubit, it is crucial to study its sensitivity to various noise sources. The qubit can decay to its ground state as a result of its interactions with the environment. The sensitivity to this relaxation process is quantified by the $T_1$ time.
We note that an undesired excitation of the qubit state may also result from such interactions.
Table~\ref{tab:T_1} provides an overview of the noise contributions that can be estimated with CircuitQ.
We included relaxation due to quasiparticle tunneling, dielectric loss and flux noise. 
\paragraph{Quasiparticle tunneling}
In experiments, significant non-vanishing densities of unpaired electrons could be observed, which are referred to as quasiparticles in this context~\cite{martinis_energy_2009}. 
Tunneling of such quasiparticles through the junction barrier can lead to a relaxation of the qubit.
This effect is separated into two contributions as given in equation~\ref{eq:T1_qp}. The first contribution concerns the junctions in the circuit (sum over $j$), while the second contribution is associated with every linear inductance in the circuit (sum over $l$) \cite{catelani_relaxation_2011}. 
The implementation of the $\sin\left( \frac{\hat{\varphi}}{2}\right) = \sin\left( \frac{\hat{\Phi}}{2\Phi_0}\right)$ operator in equation~\ref{eq:T1_qp} is straightforward in the flux basis, where it is represented by a diagonal matrix with the values of the sine function on the diagonal.
However, a more elaborate implementation is needed for the charge basis~\cite{serniak_nonequilibrium_2019}, as the $\sin\left( \frac{\hat{\varphi}}{2}\right)$ operator describes a tunneling process in the basis of single elementary charges, which can not be represented in the conventional charge basis of Cooper pairs:
\begin{align}
    \sin\left( \frac{\hat{\Phi}}{2\Phi_0}\right) \rightarrow \frac{1}{2i} \left( \sum_{\tilde{n}} \ket{\tilde{n}-1}\bra{\tilde{n}} - \ket{\tilde{n}} \bra{\tilde{n} -1} \right),
\end{align}
with $\ket{\tilde{n}}$ being the basis state of the single electron charge basis.
The implementation of this operator in the single charge basis follows the same procedure as for the cosine operator in the Cooper pair basis outlined in equations \ref{eq:cos_in_cooper_pair_basis}-\ref{eq:cos_implementation}.
However, the operator implemented in the single charge basis has dimension $2d-1$, where $d$ is the number of states in the Cooper pair charge basis. 
To calculate the transition element defined as
\begin{align}\label{eq:transition_element_qp_tunneling}
M_{eg} := \langle g| \sin \left( \frac{\hat{\Phi}_{e_{j}}}{2\Phi_0}\right) | e \rangle,
\end{align}
we have to transform the ground and excited states $\ket{g}$ and $\ket{e}$ from the Cooper pair basis to the single charge basis.
We can distinguish two different configurations of the state vectors in the single charge basis: One with only even numbered entries of the single charge basis non-zero, and the other with only the odd numbered entries occupied.
To transform from the Cooper pair basis to either the even or odd configuration, we define the respective $(2d-1) \times d $ dimensional transformation matrices $\hat{T}_{\textrm{ce}}$ and $\hat{T}_{\textrm{co}}$ with 
\begin{align}
    \hat{T}_{\textrm{ce}} =    \left( 
        \begin{matrix}
            1 & 0 & 0 & \\
            0 & 0 & 0 & \\
            0 & 1 & 0 &  \\
            0 & 0 & 0 & \cdots \\
            0 & 0 & 1 & \\
            0 & 0 & 0 & \\
              & \vdots & & 
        \end{matrix}
                \right), \ 
    \hat{T}_{\textrm{co}} =    \left( 
        \begin{matrix}
            0 & 0 & 0 & \\
            1 & 0 & 0 & \\
            0 & 0 & 0 &  \\
            0 & 1 & 0 & \cdots \\
            0 & 0 & 0 & \\
            0 & 0 & 1 & \\
              & \vdots & & 
        \end{matrix}
                \right).
\end{align}
Finally, to evaluate the transition element $M_{eg}$ in equation~\ref{eq:transition_element_qp_tunneling} in the charge basis, we compute
\begin{align}\label{eq:transition_element_qp_tunneling_imp}
M_{eg} \rightarrow \langle g| \hat{T}_{\textrm{ce}}^\dagger \sin \left( \frac{\hat{\Phi}_{e_{j}}}{2\Phi_0}\right) \hat{T}_{\textrm{co}}| e \rangle.
\end{align}

\paragraph{Dielectric Loss}
Another noise channel present in superconducting qubits is relaxation due to the fact that the electrical field, which stores capacitive energy, couples to charged fluctuators~\cite{krantz_quantum_2019}. The resulting effect on the $T_1$ time is calculated with equation~\ref{eq:T1_diel}. Here we sum over all capacitors in the circuit.

\paragraph{Flux noise}
The fluctuation of spins on the superconducting material are suspected to be the origin of flux noise~~\cite{krantz_quantum_2019}. Those fluctuations perturb the magnetic field, which stores inductive energy, effectively leading to fluctuations of the electrical current of the inductive elements. 
We estimate the corresponding contribution to the relaxation time with equation~\ref{eq:T1_flux}.
As described in section~\ref{sec:symbolic_Hamiltonian}, we assign an external flux to a subset of edges $\mathcal{B}_c$ of the circuit graph.
We therefore sum over all such edges to estimate the flux noise.
If the associated element is a Josephson junction, the current operator is given by $\hat{I}=I_C \sin \frac{\hat{\Phi}}{\Phi_0}$, while the expression $\hat{I}= \frac{\hat{\Phi}}{L}$ is used for linear inductors.
We included the possibility to obtain a lower bound on the $T_1$ estimate by summing not only over the edges in $\mathcal{B}_c$ but including all inductive edges.

The focus of this toolbox is on pure qubit design without considering qubit control such as state preparation. Therefore, we do not consider noise due to the Purcell effect for now. We also did not include pure dephasing mechanisms explicitly, which can be attributed to the fluctuation of the qubit frequency due to various noise channels. Estimating those dephasing processes would entail the calculation of the derivative of the qubit frequency with respect to the particular noise source. This derivative could be either calculated numerically or even symbolically, depending on the efficiency of those approaches. Adding dephasing processes to the analysis is an important part of our outlook, as it is an essential part of an extensive and general study of superconducting circuits.

In comparison to the related open-source software toolbox scQubits~\cite{groszkowski_scqubits_2021}, we follow a similar strategy by estimating the coherence times using Fermi's golden rule combined with the specific noise spectral densities from literature. However, our expressions for the noise spectral densities differ in the case of noise due to quasiparticle tunneling and flux noise, where we follow Nguyen et al.'s study of the fluxonium qubit~\cite{nguyen_high-coherence_2019}. We also add the contribution due to linear inductances to calculate noise due to quasiparticle tunneling.

\begin{figure*}[htb]
\includegraphics{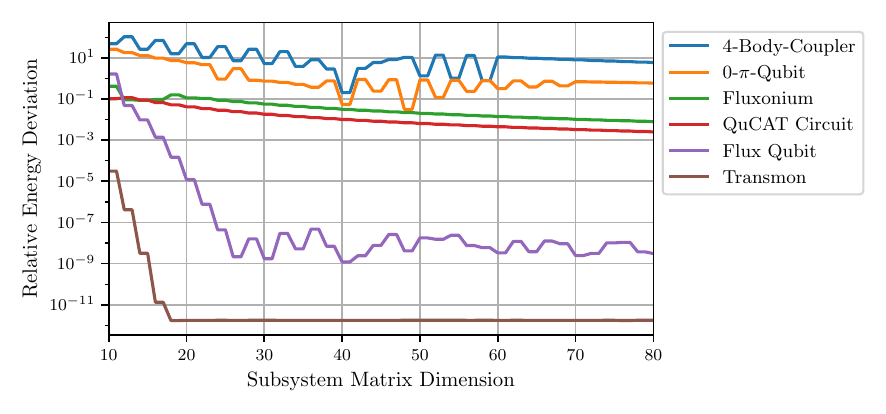}
\caption{Benchmark tests for six different circuits. We calculated the spectrum for those instances with reference implementations (see main text) and compared the transition energy between the ground state and excited state to the corresponding output of CircuitQ. The deviation is scaled by the qubit energy of the respective reference implementation. The test has been repeated for various dimensions of the subsystem matrices, which are the numerical representation of the node fluxes and charges. We note that the total dimension scales exponentially with the number of ungrounded nodes, which leads to longer evaluation times for larger circuits.}
\label{fig:benchmark}
\end{figure*}

\section{Demonstration and benchmark}\label{sec:demonstration}


To demonstrate the capabilities of CircuitQ, we use three well known circuits from the literature, i.e., the fixed-frequency transmon~\cite{koch_charge-insensitive_2007}, the fluxonium~\cite{manucharyan_fluxonium_2009} and the persistent-current flux qubit~\cite{orlando_superconducting_1999}.
The examples are initialized with the corresponding input graph, from which CircuitQ computes the symbolic and numerical Hamiltonian. The latter can be diagonalized to analyze the spectrum and eigenstates of the system.
Figure~\ref{fig:examples} gives an overview of this procedure while figure~\ref{fig:t1} depicts the $T_1$ contributions that are estimated by the toolbox for these circuits. For the noise estimates, we considered all three depolarization channels introduced in section~\ref{subsec:features}. For the fixed frequency transmon circuit, we consider two noise channels: quasiparticle tunneling and dielectric loss. For the chosen circuit parameters, the second contribution is observed to be the limiting factor. The lifetime of planar (2D) transmon fabrications are reported to be limited by dielectric loss~\cite{place_new_2021}.
As we simulate the fluxonium at the sweet spot $\tilde{\Phi}_{\textrm{ext}} = \pi \cdot \Phi_0$ here, the quasiparticle noise of the small junction (first term in equation~\ref{eq:T1_qp}) is suppressed and we can ascribe the $T_1$ decay mostly to the linear inductance (second term in equation~\ref{eq:T1_qp}). 
Similar to the transmon, the flux qubit is limited by the dielectric loss when comparing the three relaxation processes. The parameter values used for estimating the lifetime of a flux qubit refer to \textit{Qubit B} in reference~\cite{yan_flux_2016}, where the lifetime at the sweet spot $\tilde{\Phi}_{\textrm{ext}} = \pi \cdot \Phi_0$ seems to be limited by flux noise. Our findings indicate that our specifications chosen for the flux and charge noise estimates do not resemble this particular experimental set-up accurately. However, our estimate for the effective $T_1$ time lies in the same order of magnitude compared to the findings in this reference.   
The effective $T_1$ time for the transmon and fluxonium range in between $10^2-10^3$ $\mu$s, while this value is reduced by one order of magnitude for the flux qubit.
These numbers are in accordance with values from literature~\cite{kjaergaard_superconducting_2020}.
We note that the exact estimate of the $T_1$ time depends on parameters like the density of quasiparticles or the dielectric quality factor. Those parameters depend on the specific realization of the circuits and will vary from experiment to experiment.
Although we have chosen representative values, the computed $T_1$ times should not be understood to be exact for a specific circuit layout but should serve as estimates to classify the sensitivity of a circuit to certain noise channels.

In Appendix~\ref{appendix:code_fig_examples}, code samples are provided to initialize the instances for the example circuits. 
The symbolic Hamiltonian that is generated for the transmon by the toolbox is
\begin{align}
    H = - E_{J010} \cos{\left( \frac{\Phi_{1}}{\Phi_0} \right)} + \frac{\left(q_{1} + \tilde{q}_{1} \right)^{2}}{2C_{01}}.
\end{align}
Here, $\Phi_{1}$ and $q_{1}$ are the flux and charge variables of node $1$, and $\tilde{q}_{1}$ is a charge offset that can be introduced upon initialization (see Appendix~\ref{appendix:code_fig_examples}).
We associate $E_{J010}$ with the Josephson energy of the $0$-th junction between node $0$ and $1$, which is shunted by the capacitance $C_{01}$.
This notation, which assigns circuit elements like a Josephson junction to Hamiltonian parameters like a Joesphson energy by providing the corresponding edge nodes in the index of the symbols, is kept consistent.
For better readability, we do not define all the symbols in the following Hamiltonians individually.
The flux quantum $\Phi_0$ will be displayed as $\Phi_o$ in the toolbox, to distinguish it from the node flux of node $0$.
The numerical values for the corresponding spectrum plot in figure~\ref{fig:examples}, which shows the lowest eigenstates of the weakly anharmonic cosine-potential, are the default values, i.e. $C_{01}=100 $~fF and $E_{J010} \approx 9.69~$GHz$\cdot h$.

For the fluxonium qubit, the toolbox determines the symbolic Hamiltonian 
\begin{align}
    H = - E_{J010} \cos{\left( \frac{\Phi_{1}}{\Phi_0} \right)} + \frac{\left(\Phi_{1} + \tilde{\Phi}_{010}\right)^{2}}{2 L_{010}} + \frac{q_{1}^{2}}{2C_{01}},
\end{align}
where $\tilde{\Phi}_{010}$ labels the offset flux of the inductive loop. An excerpt of the spectrum of this Hamiltonian is shown in figure~\ref{fig:examples} with $C_{01}=10$~fF, $L_{010}=0.5~\mu$H, $E_{J010}~\approx~48.43~$GHz$\cdot h$ and $\tilde{\Phi}_{010} = \pi \Phi_0$. It shows the typical low-lying 0 and 1 states that are localized in the wells, with higher plasma states several GHz above.

Finally, the persistent-current flux qubit Hamiltonian constructed by the toolbox can be written as
\begin{align}
    H =& - E_{J010} \cos{\left( \frac{\Phi_{1}}{\Phi_0} \right)} - E_{J020} \cos{\left( \frac{\Phi_{2}}{\Phi_0} \right)} \nonumber \\
    &- E_{J120} \cos{\left( \frac{\Phi_{2}-\Phi_{1}+\tilde{\Phi}_{120}}{\Phi_0} \right)} \nonumber \\
    &+ \frac{q_{1}^2 \left(C_{02} + C_{12} \right) + 2 q_1 q_2 C_{12} + q_{2}^2 \left(C_{01} + C_{12} \right)}{2 \left( C_{01}C_{02} + C_{01}C_{12} + C_{02}C_{12} \right)}.
\end{align}
We depict the ground state of this Hamiltonian in figure~\ref{fig:examples} for $\alpha = 0.7$, $C_{01}=C_{02}=\frac{C_{12}}{\alpha}=50$~fF, $E_{J010}=E_{J020}=\frac{E_{J020}}{\alpha} \approx 9.69~$GHz$\cdot h$ and $\tilde{\Phi}_{020} = \pi \Phi_0$.
The ground state is localized in the double well potential, which is repeated periodically throughout the chosen flux grid.
As for the transmon circuit, due to the periodicity of the potential, the Hamiltonian is implemented in the charge basis. For figure~\ref{fig:examples}, we use the transformation method of the toolbox to visualize the eigenstates in the flux basis.


In order to test the software and to check the accuracy of our numerical implementation, we perform benchmark tests for a variety of circuits.
In addition to the three example circuits that have been discussed in this section, we complete the benchmark by adding the 0-$\pi$-Qubit~\cite{brooks_protected_2013}, the 4-body-coupler which is referred to as \textit{Circuit C} in reference~\cite{menke_automated_2021} and the circuit of a transmon that is capacitively coupled to a resonator to the list of test circuits.
The latter circuit is called the QuCAT circuit here, as a similar version is discussed in the corresponding reference~\cite{gely_qucat_2020}.
The code to construct the CircuitQ instances can be found in Appendix~\ref{appendix:code_benchmark_circuits}.
We use existing software implementations to calculate the spectrum of the test circuits as a benchmark and compare the results of CircuitQ to it.
As a reference, we used the toolbox scQubits~\cite{groszkowski_scqubits_2021} for the transmon, fluxonium, 0-$\pi$-qubit and persistent-current flux qubit circuit.
The QuCAT circuit has been compared to its implementation in the QuCAT toolbox~\cite{gely_qucat_2020}.
The 4-body-coupler has been tested against a direct and individualized software implementation.
As the test outcome depends on the size of the numerical matrices which represent the charge and flux variables, we vary the dimension of those matrices.
The result is shown in figure~\ref{fig:benchmark}.
CircuitQ automatically implements the transmon and flux qubit in the charge basis, and the fluxonium,  0-$\pi$-qubit and 4-body-coupler in the flux basis.
The QuCAT circuit is implemented in a mixture of both bases.
For the transmon, fluxonium and flux qubit as well as for the QuCAT circuit, we find a good agreement between CircuitQ and the benchmark implementation. As detailed in appendix~\ref{app:fluxonium_limitations}, it is still possible to observe numerical limitations on less complex circuits like the fluxonium qubit.
We observe larger deviations for the 4-body-coupler and the 0-$\pi$-qubit, which are more complex circuits, even for large numerical matrices.
For some circuits, a quantization of the node variables is not the most natural choice, as characteristic modes of the system might be a combination of several node variables.
To find a more natural quantization, a coordinate transformation of the node variables can be performed prior to quantization.
Therefore, the deviation for the 4-body-coupler and the 0-$\pi$-qubit can be attributed to the lack of an appropriate coordinate transformation.  

\section{Conclusion}
We presented the core functionalities of CircuitQ. With the ability to derive a symbolic and numerical Hamiltonian from a superconducting circuit in an automated way, CircuitQ can serve as a toolbox for the community to analyze superconducting circuits. The input circuit can be a general superconducting circuit that combines Josephson junctions, linear inductances and capacitances. An automated procedure to analyze superconducting circuits is a benefitial tool for the study of superconducting circuits within the context of quantum information. Apart from the application to computing, superconducting circuits can be also used as a platform in other areas of application like sensing~\cite{danilin_quantum_2021} or studying thermodynamics~\cite{kerremans_probabilistically_2022}. 

While the toolbox is currently limited to the computation of few-node circuits, future work should address the optimization of speed and scalability. As an example, on a conventional personal computer, it took below $1$\,s to initialize an instance of the transmon circuit and calculate the lowest $10$ eigenstates and eigenvalues of the numerical Hamiltonian for subsystem matrix dimensions $40$ and $80$, while for the 0-$\pi$-Qubit, it took around $72$\,s for a subsystem dimension $40$ and around $105$ s for a subsystem dimension $80$, still with lacking accuracy as described in the section~\ref{sec:demonstration}. A key feature of CircuitQ is its dynamic implementation in the charge and flux basis. At the moment, the variables that are quantized are always the node variables of the circuit graph. For some circuits, it is crucial to perform a variable transformation prior to quantization. Adding a suitable transformation would represent an important development step towards the goal of increasing the calculation speed.
Another improvement can be made by implementing hierarchical diagonalization such as discussed in reference~\cite{kerman_efficient_2020}.
In addition, the toolbox is written in a modular fashion that allows for extensions towards time-dependent simulations.

Thanks to the general functionality, the possibility to include charge and flux offsets, as well as the the incorporation of noise estimates, CircuitQ can serve as a versatile tool for the design of superconducting qubits. Adding more noise channels, especially estimates for the dephasing time, would be an important future addition to the software. Moreover, adding the possibility to incorporate external impedances as circuit elements would allow to estimate noise from first principles~\cite{burkard_multilevel_2004}.

\section*{Acknowledgements}
We are thankful for fruitful discussions with Jens Koch and his group as well as with Kyle Serniak and Andrew J. Kerman. We also thank three anonymous reviewers for their helpful suggestions. PA thanks Glen Bigan Mbeng, Kilian Ender and Benoît Vermersch for helpful discussions and Martin Lanthaler for designing the logo. This work was supported by the Austrian Science Fund (FWF) through a START grant under Project No. Y1067-N27 and the SFB BeyondC Project No. F7108-N38, the Hauser-Raspe foundation, and the European Union's Horizon 2020 research and innovation program under grant agreement No. 817482. This material is based upon work supported by the Defense Advanced Research Projects Agency (DARPA) under Contract No. HR001120C0068. Any opinions, findings and conclusions or recommendations expressed in this material are those of the authors and do not necessarily reflect the views of DARPA.
TM acknowledges funding by the Office of the Director of National Intelligence (ODNI), Intelligence Advanced Research Projects Activity (IARPA) under Air Force Contract No. FA8721-05-C-0002.
The views and conclusions contained herein are those of the authors and should not be interpreted as necessarily representing the official policies or endorsements, either expressed or implied, of the ODNI, IARPA, or the U.S. Government.
The U.S. Government is authorized to reproduce and distribute reprints for Governmental purposes notwithstanding any copyright annotation thereon.

\newpage

\clearpage
\newpage

\appendix

\section{Parameter values for $T_1$ estimates.}\label{appendix:values_T1}
Figure~\ref{fig:t1} in the main text provides an overview of the $T_1$ estimates provided by CircuitQ for the example circuits studied in this article, i.e. transmon, fluxonium and flux qubit. In table~\ref{tab:values_T1}, we list the parameter values chosen for the purpose of this illustration. The numbers for the fluxonium correspond to \textit{Qubit A} from reference~\cite{nguyen_high-coherence_2019}, while the values for the flux qubit are associated to \textit{Qubit B} from reference~\cite{yan_flux_2016}.
\begin{table}[ht]
\centering
\caption{Overview of the parameter values used to calculate the $T_1$ times shown in figure~\ref{fig:t1} for three example circuits. The energies given for the flux qubit relate to the large junctions and the values have to be scaled by $\alpha$ to deduce the corresponding numbers for the small junction. Both the fluxonium and flux qubit are evaluated at the sweet spot. The energy values are given in frequencies and have to be scaled by $h$ to obtain units of energy. If a value is calculated, it is displayed as rounded to two decimal places. }
\label{tab:values_T1}
\begin{tabular}{p{1.7cm} p{1.7cm} p{1.7cm} p{1.7cm} p{1.3cm}}
	\hline 
Circuit	& Parameters & &  & \\ 
\hline \hline
Transmon & $E_J$ & $E_C$  & & \\
 & 10 GHz & 0.24 GHz & & \\
\hline 
Fluxonium & $E_J$ & $E_C$ & $E_L$ & $\tilde{\Phi}_{\textrm{ext}}$   \\
 & 3 GHz & 0.8 GHz&  1 GHz & $\pi \cdot \Phi_0$ \\
\hline 
Flux Qubit & $E_J$ & $E_C$ & $\alpha$ & $\tilde{\Phi}_{\textrm{ext}}$ \\
 & 86.19 GHz & 0.15 GHz & 0.42 & $\pi \cdot \Phi_0$  \\
\hline 
\end{tabular} 
\end{table}

\section{Code samples for illustrative circuits in figure~\ref{fig:examples} }\label{appendix:code_fig_examples}

In the following subsections, we present the code that generates the three instances which are displayed in figure~\ref{fig:examples} and which are discussed in Sec~\ref{sec:demonstration}.

\subsection{Transmon}

\begin{lstlisting}[language=Python]
import circuitq as cq
import networkx as nx

graph = nx.MultiGraph()
graph.add_edge(0,1, element = 'C')
graph.add_edge(0,1, element = 'J')
circuit = cq.CircuitQ(graph, offset_nodes=[1])

# Numerical implementation and diagonalisation
h_num = circuit.get_numerical_hamiltonian(400,
              grid_length=np.pi*circuit.phi_0)
eigv, eigs = circuit.get_eigensystem()

# Conversion for the plot
circuit.transform_charge_to_flux()
eigs = circuit.estates_in_phi_basis
\end{lstlisting}

\subsection{Fluxonium}
\begin{lstlisting}[language=Python]
import circuitq as cq
import networkx as nx

graph = nx.MultiGraph()
graph.add_edge(0,1, element = 'C')
graph.add_edge(0,1, element = 'J')
graph.add_edge(0,1, element = 'L')
circuit = cq.CircuitQ(graph)

# Numerical implementation and diagonalisation
EJ = circuit.c_v["E"]*.5
L = circuit.c_v["L"]*5
C = circuit.c_v["C"]*0.1
phi_ext =np.pi*circuit.phi_0 
h_num = circuit.get_numerical_hamiltonian(400, 
        parameter_values=[C, EJ, L, phi_ext])
eigv, eigs = circuit.get_eigensystem()
\end{lstlisting}

\subsection{Persistent-current flux qubit}
\begin{lstlisting}[language=Python]
import circuitq as cq
import networkx as nx
 
graph = nx.MultiGraph()
graph.add_edge(0,1, element = 'C')
graph.add_edge(0,1, element = 'J')
graph.add_edge(1,2, element = 'C')
graph.add_edge(1,2, element = 'J')
graph.add_edge(0,2, element = 'C')
graph.add_edge(0,2, element = 'J')
circuit = cq.CircuitQ(graph)

# Numerical implementation and diagonalisation
dim = 51
EJ = 1*circuit.c_v["E"]
alpha = 0.7
C = circuit.c_v["C"]*0.5
phi_ext = np.pi*circuit.phi_0 
h_num = circuit.get_numerical_hamiltonian(dim,
parameter_values=[C,C,alpha*C,
                  EJ,EJ,alpha*EJ,phi_ext])
eigv, eigs = circuit.get_eigensystem()

# Conversion for the plot
circuit.transform_charge_to_flux()
eigs = circuit.estates_in_phi_basis

\end{lstlisting}

\section{Code sample for example circuit in figure~\ref{fig:circuit}}\label{appendix:code_fig_circuit}
The following code demonstrates the initialization of the circuit in figure~\ref{fig:circuit}.
\begin{lstlisting}[language=Python]
import circuitq as cq
import networkx as nx

graph = nx.MultiGraph()
graph.add_edge(0,1, element = 'C')
graph.add_edge(0,2, element = 'J')
graph.add_edge(0,2, element = 'C')
graph.add_edge(1,2, element = 'L')
graph.add_edge(1,2, element = 'C')
graph.add_edge(1,3, element = 'C')
graph.add_edge(1,3, element = 'L')
graph.add_edge(2,3, element = 'L')
graph.add_edge(2,3, element = 'C')

circuit = cq.CircuitQ(graph)
\end{lstlisting}

\section{Code sample for additional benchmark circuits} \label{appendix:code_benchmark_circuits}
In the following, we present the generation of the numerical Hamiltonian for the additional benchmark circuits in figure~\ref{fig:benchmark}, which have not been given in Appendix~\ref{appendix:code_fig_examples} yet, for an arbitrary subsystem matrix dimension $d$. We note that for this benchmark task, contrary to the parameter values given in the code samples in Appendix~\ref{appendix:code_fig_examples}, we choose the values given in table~\ref{tab:values_T1} for the listed circuits.

\subsection{0-$\pi$-Qubit}

\begin{lstlisting}[language=Python]
import circuitq as cq
import networkx as nx

graph = nx.MultiGraph()
graph.add_edge(1,2, element = 'C')
graph.add_edge(1,2, element = 'J')
graph.add_edge(2,3, element = 'L')
graph.add_edge(3,4, element = 'J')
graph.add_edge(3,4, element = 'C')
graph.add_edge(4,1, element = 'L')
graph.add_edge(1,3, element = 'C')
graph.add_edge(2,4, element = 'C')

circuit = cq.CircuitQ(graph, ground_nodes=[1])
h_num = circuit.get_numerical_hamiltonian(d,
    parameter_values=   [False,
                        100 * circuit.c_v['C'],
                        100 * circuit.c_v['C'],
                        False, False,
                        False, False,
                        False, False,
                        False, False]    )
\end{lstlisting}

\subsection{4-Body-Coupler}

\begin{lstlisting}[language=Python]
import circuitq as cq
import networkx as nx
import numpy as np

graph = nx.MultiGraph()
graph.add_edge(0,2, element = 'C')
graph.add_edge(0,2, element = 'J')
graph.add_edge(2,3, element = 'L')
graph.add_edge(2,3, element = 'C')
graph.add_edge(0,3, element = 'C')
graph.add_edge(0,3, element = 'J')
graph.add_edge(2,1, element = 'C')
graph.add_edge(2,1, element = 'J')
graph.add_edge(1,3, element = 'L')
graph.add_edge(1,3, element = 'C')

circuit = cq.CircuitQ(graph, ground_nodes=[0])
L13 = 289.395  # in pH
L23 = 120.416  # in pH
lj12 = 3.75498  # in um
lj22 = 0.395517  # in um
lj33 = 0.373288  # in um
Jc = 5e-6  # critical current density in A/um^2
wJ = 0.2  # junction width in um
Sc = 60e-15  # specific capacitance in F/um^2
Phi0 = 2.06783385 * 10 ** (-15)  # flux quantum
L13 = L13 * 1e-12  # scale pH -> H
L23 = L23 * 1e-12  # scale pH -> H
Ej12 = Phi0 / (2 * np.pi) * Jc * wJ * lj12
Ej22 = Phi0 / (2 * np.pi) * Jc * wJ * lj22
Ej33 = Phi0 / (2 * np.pi) * Jc * wJ * lj33
circuit.get_numerical_hamiltonian(d,
parameter_values=
        [4.455 * 1e-15 + Sc * wJ * lj22, 
        70.556 * 1e-15 + Sc * wJ * lj33,
        Sc * wJ * lj12, 
        16.832 * 1e-15, 
        85.677 * 1e-15,
        Ej22, Ej33, Ej12, L23, L13, 
        0.5 * circuit.phi_0, 
        0.02 * circuit.phi_0]     )
\end{lstlisting}

\subsection{QuCAT circuit}

\begin{lstlisting}[language=Python]
import circuitq as cq
import networkx as nx
import numpy as np

graph = nx.MultiGraph()
graph.add_edge(0,1, element = 'C')
graph.add_edge(0,1, element = 'J')
graph.add_edge(0,2, element = 'C')
graph.add_edge(0,2, element = 'L')
graph.add_edge(1,2, element = 'C')

circuit = cq.CircuitQ(graph, ground_nodes=[0])
circuit.get_numerical_hamiltonian(d,                 
            grid_length=np.pi*circuit.phi_0,
            parameter_values=[100e-15,100e-15,
                1e-15,(circuit.phi_0**2)/8e-9, 
                10e-9 ]          )
\end{lstlisting}

\section{Symbolic Hamiltonian for circuit in figure~\ref{fig:circuit}}\label{appendix:leib_circuit_hamiltonian}
CircuitQ provides the symbolic Hamiltonian given in equation~\ref{eq:leib_circuit} for the example circuit in figure~\ref{fig:circuit}. The parasitic capacitances within the kinetic part of the Hamiltonian are labelled with a $p$ in the index.

\begin{widetext}
\begin{align}
H = & \left( C_{01} C_{02} C_{13} + C_{01} C_{02} Cp_{23} + C_{01} C_{13} Cp_{12} + C_{01} C_{13} Cp_{23} \right. \nonumber \\ 
& \left. \ + C_{01} Cp_{12} Cp_{23} + C_{02} C_{13} Cp_{12} + C_{02} C_{13} Cp_{23} + C_{02} Cp_{12} Cp_{23} \right)^{-1} \cdot \nonumber \\
&  \quad \cdot \bigg(  \frac{q_{1}}{2} \Big( q_{1} \left(C_{02} C_{13} + C_{02} Cp_{23} + C_{13} Cp_{12} + C_{13} Cp_{23} + Cp_{12} Cp_{23}\right) \nonumber \\
& \qquad \qquad + q_{2} \left(C_{13} Cp_{12} + C_{13} Cp_{23} + Cp_{12} Cp_{23}\right)  \nonumber \\
& \qquad \qquad + q_{3} \left(C_{02} C_{13} + C_{13} Cp_{12} + C_{13} Cp_{23} + Cp_{12} Cp_{23}\right) \Big) \nonumber \\
&  \qquad + \frac{q_{2}}{2} \Big( q_{1} \left(C_{13} Cp_{12} + C_{13} Cp_{23} + Cp_{12} Cp_{23}\right) \nonumber \\
& \qquad \qquad + q_{2} \left(C_{01} C_{13} + C_{01} Cp_{23} + C_{13} Cp_{12} + C_{13} Cp_{23} + Cp_{12} Cp_{23}\right) \nonumber \\
& \qquad \qquad + q_{3} \left(C_{01} Cp_{23} + C_{13} Cp_{12} + C_{13} Cp_{23} + Cp_{12} Cp_{23}\right)\Big) \nonumber \\
& \qquad + \frac{q_{3}}{2} \Big(q_{1} \left(C_{02} C_{13} + C_{13} Cp_{12} + C_{13} Cp_{23} + Cp_{12} Cp_{23}\right) \nonumber \\ 
& \qquad \qquad + q_{2} \left(C_{01} Cp_{23} + C_{13} Cp_{12} + C_{13} Cp_{23} + Cp_{12} Cp_{23}\right) \nonumber \\
& \qquad \qquad + q_{3} \left(C_{01} C_{02} + C_{01} Cp_{12} + C_{01} Cp_{23} + C_{02} C_{13} + C_{02} Cp_{12} + C_{13} Cp_{12} + C_{13} Cp_{23} + Cp_{12} Cp_{23}\right)\Big) \bigg) \nonumber \\
& - E_{J020} \cos{\left( \frac{\Phi_{2}}{\Phi_0} \right)} - E_{J130} \cos{\left( \frac{\Phi_{1} - \Phi_{3}}{\Phi_0} \right)} \nonumber \\
&  + \frac{\left(\Phi_{3} - \Phi_{2} + \tilde{\Phi}_{230}\right)^{2}}{2 L_{230}} + \frac{\left(\Phi_{2} - \Phi_{1} \right)^{2}}{2 L_{120}} \label{eq:leib_circuit}
\end{align}
\end{widetext}

\section{Limitations of the numerical treatment}\label{app:fluxonium_limitations}
\begin{figure*}[htb]
\includegraphics{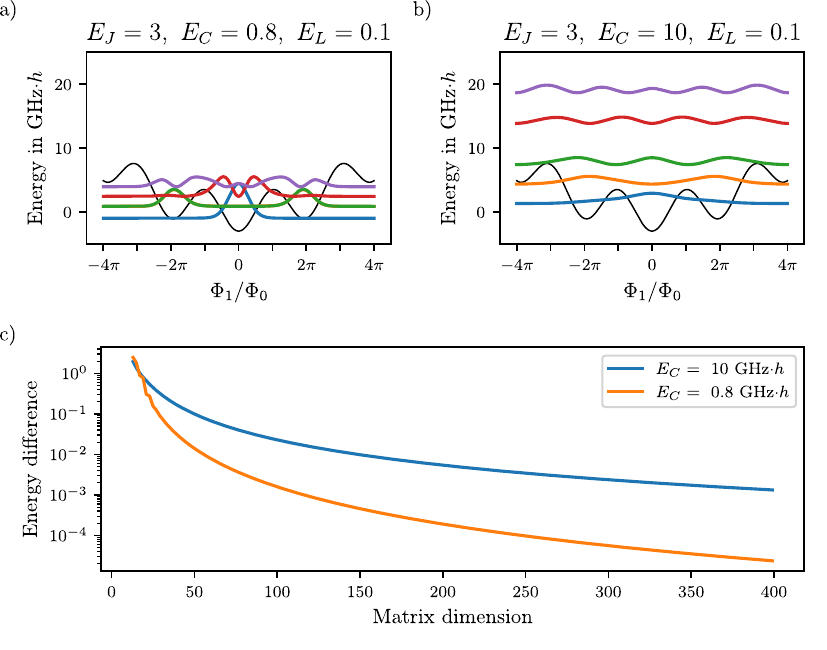}
\caption{Numerical limitations tested on the fluxonium circuit. (a), (b) Depiction of the lower energy spectrum of the fluxonium circuit for low and high $E_C$ in the same fashion as in the last row of figure~\ref{fig:examples}. The numerical parameter values chosen are given in the title of the subfigures, while $\tilde{\Phi}_{\text{ext}}=0$ for both subfigures. The black line indicates the potential energy. (c) Difference of the eigenenergies corresponding to the instances introduced in (a) and (b) as a function of the dimension of the numerical matrix representing the numerical Hamiltonian. To calculate the energy difference for a given matrix dimension we sum over the differences between the lowest $5$ eigenvalues and the eigenvalues calculated for the previous matrix dimension. The energy difference is in units of GHz$\cdot h$ and scaled logarithmically. Here, we solely use odd numbers as matrix dimension, as even numbers for the dimension will be converted to odd numbers by CircuitQ.}
\label{fig:fluxonium_deviation}
\end{figure*}
\begin{figure*}[htb]
\includegraphics{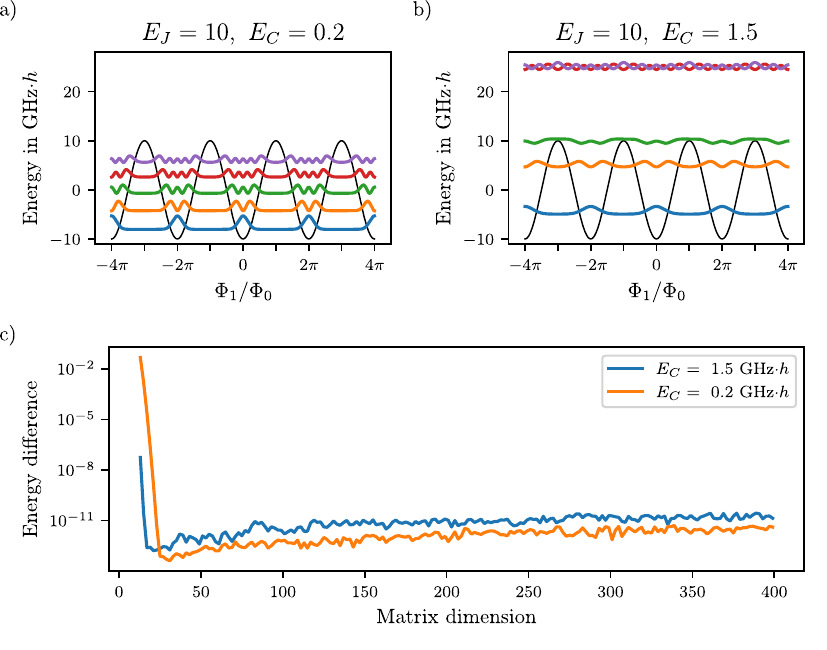}
\caption{Numerical limitations tested on the transmon circuit. (a), (b) Lower energy spectrum of the transmon circuit for low and higher $E_C$ depicted in the same fashion as in figure \ref{fig:fluxonium_deviation}(a) and \ref{fig:fluxonium_deviation}(b). Due to the charge basis implementation, to illustrate the spectrum as a function of flux, the eigenstates are transformed to the flux basis by the basis transformation outlined in section~\ref{subsubsec:charge_basis}. (c) Difference of the eigenenergies corresponding to the instances introduced in (a) and (b) depicted in the same fashion as in figure \ref{fig:fluxonium_deviation}(c).}
\label{fig:transmon_deviation}
\end{figure*}
As discussed in section~\ref{sec:demonstration}, the limitations of our numerical implementation become evident for complex circuits like the 0-$\pi$-Qubit. However, it is possible to investigate the numerical limitations on more simple circuits like the fluxonium qubit. In figure~\ref{fig:fluxonium_deviation}(a) and \ref{fig:fluxonium_deviation}(b), we depict the lowest eigenstates together with the potential energy of the fluxonium qubit for low and high capacitive energy $E_C$. The inductive energy $E_L$ has been kept small to avoid strong confinement. We observe the lowest eigenstates to be located within the potential wells for the case of low $E_C$, while the eigenstates for higher $E_C$ tend to become more delocalized. In CircuitQ, the fluxonium circuit will be implemented in the flux basis, which works well for localized states that are trapped in a harmonic potential. To measure the numerical accuracy of the software implementation, figure~\ref{fig:fluxonium_deviation}(c) shows the deviation of the eigenenergies for the instances of subfigure \ref{fig:fluxonium_deviation}(a) and \ref{fig:fluxonium_deviation}(b) as a function of the numerical matrix dimension. In an ideal case, the spectrum is almost invariant under slight changes of the matrix dimension. Such a convergence can be observed for high values of matrix dimension. However, for lower values of matrix dimension, we observe a significant deviation of the energy, which is, besides the regime of very low matrix dimension, drastically higher for the case of high $E_C$. This indicates that for the case of weakly localized wavefunctions, the flux basis implementation of the toolbox is reaching its numerical limitation. Moreover, another inaccuracy is introduced for higher lying states due to the cut-off of the numerical flux grid from $-4\pi$ to $4 \pi$.

In figure~\ref{fig:transmon_deviation}, we present a similar study for the fixed-frequency transmon circuit (see figure~\ref{fig:examples}(a)), which will be implemented in the charge basis. Figure~\ref{fig:transmon_deviation}(a) depicts the spectrum of the transmon circuit for low $E_C$ with $E_J/E_C = 50$. Here, the wavefunctions are periodically localized within the potential wells. In figure~\ref{fig:transmon_deviation}(b) we show the spectrum for higher $E_C$ with $E_J/E_C \approx 7$, where higher eingestates become less confined. As before, we depict the difference in energy as a function of matrix dimension in figure~\ref{fig:transmon_deviation}(c). The energy difference drops of fast for both instances and fluctuates due to numerical fluctuations at small values for high matrix dimension. In comparison to the fluxonium study in figure~\ref{fig:fluxonium_deviation}(c), here, the depicted values of energy difference are small as the $y$-axis is scaled to smaller values. This indicates, that, in contrast to the flux basis, the charge basis is well suited to describe delocalized states. The states lying energetically above the potential can be associated with free particles, which, in the case of conventional mechanics, are efficiently described in the momentum basis. As the charge basis is analogous to the momentum basis, we find a more accurate description for the transmon circuit with high capacitive energy in comparison to the case of the fluxonium circuit.


\begin{thebibliography}{10}
	\expandafter\ifx\csname url\endcsname\relax
	\def\url#1{\texttt{#1}}\fi
	\expandafter\ifx\csname urlprefix\endcsname\relax\def\urlprefix{URL }\fi
	\providecommand{\bibinfo}[2]{#2}
	\providecommand{\eprint}[2][]{\url{#2}}
	
	\bibitem{arute_quantum_2019}
	\bibinfo{author}{Arute, F.} \emph{et~al.}
	\newblock \bibinfo{title}{Quantum supremacy using a programmable
		superconducting processor}.
	\newblock \emph{\bibinfo{journal}{Nature}} \textbf{\bibinfo{volume}{574}},
	\bibinfo{pages}{505--510} (\bibinfo{year}{2019}).
	\newblock \urlprefix\url{https://www.nature.com/articles/s41586-019-1666-5}.
	
	\bibitem{blais2020circuit}
	\bibinfo{author}{Blais, A.}, \bibinfo{author}{Grimsmo, A.~L.},
	\bibinfo{author}{Girvin, S.} \& \bibinfo{author}{Wallraff, A.}
	\newblock \bibinfo{title}{Circuit quantum electrodynamics}.
	\newblock \emph{\bibinfo{journal}{Reviews of Modern Physics}}
	\textbf{\bibinfo{volume}{93}}, \bibinfo{pages}{025005}
	(\bibinfo{year}{2021}).
	\newblock
	\urlprefix\url{https://link.aps.org/doi/10.1103/RevModPhys.93.025005}.
	
	\bibitem{kjaergaard_superconducting_2020}
	\bibinfo{author}{Kjaergaard, M.} \emph{et~al.}
	\newblock \bibinfo{title}{Superconducting {Qubits}: {Current} {State} of
		{Play}}.
	\newblock \emph{\bibinfo{journal}{Annual Review of Condensed Matter Physics}}
	\textbf{\bibinfo{volume}{11}}, \bibinfo{pages}{369--395}
	(\bibinfo{year}{2020}).
	\newblock
	\urlprefix\url{https://www.annualreviews.org/doi/10.1146/annurev-conmatphys-031119-050605}.
	
	\bibitem{weiss_spectrum_2019}
	\bibinfo{author}{Weiss, D.~K.}, \bibinfo{author}{Li, A. C.~Y.},
	\bibinfo{author}{Ferguson, D.~G.} \& \bibinfo{author}{Koch, J.}
	\newblock \bibinfo{title}{Spectrum and coherence properties of the
		current-mirror qubit}.
	\newblock \emph{\bibinfo{journal}{Physical Review B}}
	\textbf{\bibinfo{volume}{100}}, \bibinfo{pages}{224507}
	(\bibinfo{year}{2019}).
	\newblock \urlprefix\url{https://link.aps.org/doi/10.1103/PhysRevB.100.224507}.
	
	\bibitem{paolo_control_2019}
	\bibinfo{author}{Paolo, A.~D.}, \bibinfo{author}{Grimsmo, A.~L.},
	\bibinfo{author}{Groszkowski, P.}, \bibinfo{author}{Koch, J.} \&
	\bibinfo{author}{Blais, A.}
	\newblock \bibinfo{title}{Control and coherence time enhancement of the 0-$\pi$
		qubit}.
	\newblock \emph{\bibinfo{journal}{New Journal of Physics}}
	\textbf{\bibinfo{volume}{21}}, \bibinfo{pages}{043002}
	(\bibinfo{year}{2019}).
	\newblock \urlprefix\url{https://doi.org/10.1088%2F1367-2630%2Fab09b0}.
	
	\bibitem{nguyen_high-coherence_2019}
	\bibinfo{author}{Nguyen, L.~B.} \emph{et~al.}
	\newblock \bibinfo{title}{High-{Coherence} {Fluxonium} {Qubit}}.
	\newblock \emph{\bibinfo{journal}{Physical Review X}}
	\textbf{\bibinfo{volume}{9}}, \bibinfo{pages}{041041} (\bibinfo{year}{2019}).
	\newblock \urlprefix\url{https://link.aps.org/doi/10.1103/PhysRevX.9.041041}.
	
	\bibitem{mirrahimi_dynamically_2014}
	\bibinfo{author}{Mirrahimi, M.} \emph{et~al.}
	\newblock \bibinfo{title}{Dynamically protected cat-qubits: a new paradigm for
		universal quantum computation}.
	\newblock \emph{\bibinfo{journal}{New Journal of Physics}}
	\textbf{\bibinfo{volume}{16}}, \bibinfo{pages}{045014}
	(\bibinfo{year}{2014}).
	\newblock
	\urlprefix\url{https://iopscience.iop.org/article/10.1088/1367-2630/16/4/045014}.
	
	\bibitem{burkard_multilevel_2004}
	\bibinfo{author}{Burkard, G.}, \bibinfo{author}{Koch, R.~H.} \&
	\bibinfo{author}{DiVincenzo, D.~P.}
	\newblock \bibinfo{title}{Multilevel quantum description of decoherence in
		superconducting qubits}.
	\newblock \emph{\bibinfo{journal}{Physical Review B}}
	\textbf{\bibinfo{volume}{69}}, \bibinfo{pages}{064503}
	(\bibinfo{year}{2004}).
	\newblock \urlprefix\url{https://link.aps.org/doi/10.1103/PhysRevB.69.064503}.
	
	\bibitem{vool_introduction_2017}
	\bibinfo{author}{Vool, U.} \& \bibinfo{author}{Devoret, M.}
	\newblock \bibinfo{title}{Introduction to quantum electromagnetic circuits}.
	\newblock \emph{\bibinfo{journal}{International Journal of Circuit Theory and
			Applications}} \textbf{\bibinfo{volume}{45}}, \bibinfo{pages}{897--934}
	(\bibinfo{year}{2017}).
	\newblock
	\urlprefix\url{https://onlinelibrary.wiley.com/doi/abs/10.1002/cta.2359}.
	
	\bibitem{kerman_efficient_2020}
	\bibinfo{author}{Kerman, A.~J.}
	\newblock \bibinfo{title}{Efficient numerical simulation of complex {Josephson}
		quantum circuits}.
	\newblock \emph{\bibinfo{journal}{arXiv:2010.14929 [quant-ph]}}
	(\bibinfo{year}{2020}).
	\newblock \urlprefix\url{http://arxiv.org/abs/2010.14929}.
	
	\bibitem{groszkowski_scqubits_2021}
	\bibinfo{author}{Groszkowski, P.} \& \bibinfo{author}{Koch, J.}
	\newblock \bibinfo{title}{Scqubits: a {Python} package for superconducting
		qubits}.
	\newblock \emph{\bibinfo{journal}{Quantum}} \textbf{\bibinfo{volume}{5}},
	\bibinfo{pages}{583} (\bibinfo{year}{2021}).
	\newblock \urlprefix\url{https://quantum-journal.org/papers/q-2021-11-17-583/}.
	
	\bibitem{gely_qucat_2020}
	\bibinfo{author}{Gely, M.~F.} \& \bibinfo{author}{Steele, G.~A.}
	\newblock \bibinfo{title}{{QuCAT}: quantum circuit analyzer tool in {Python}}.
	\newblock \emph{\bibinfo{journal}{New Journal of Physics}}
	\textbf{\bibinfo{volume}{22}}, \bibinfo{pages}{013025}
	(\bibinfo{year}{2020}).
	\newblock
	\urlprefix\url{https://iopscience.iop.org/article/10.1088/1367-2630/ab60f6/meta}.
	
	\bibitem{andrey_klots_andreyklotssuperquantpackage_2021}
	\bibinfo{author}{Klots, A.}
	\newblock \bibinfo{title}{andreyklots/{SuperQuantPackage}}
	(\bibinfo{year}{2021}).
	\newblock \urlprefix\url{https://github.com/andreyklots/SuperQuantPackage}.
	
	\bibitem{qiskit_2022}
	\bibinfo{title}{Qiskit {Metal}} (\bibinfo{year}{2022}).
	\newblock \urlprefix\url{https://github.com/Qiskit/qiskit-metal}.
	
	\bibitem{noauthor_iqm-finlandkqcircuits_2022}
	\bibinfo{title}{iqm-finland/{KQCircuits}} (\bibinfo{year}{2022}).
	\newblock \urlprefix\url{https://github.com/iqm-finland/KQCircuits}.
	
	\bibitem{Note1}
	\bibinfo{note}{Link to GitHub repository: \protect \url
		{https://github.com/PhilippAumann/circuitq}}.
	
	\bibitem{girvin_circuit_2014}
	\bibinfo{author}{Girvin, S.~M.}
	\newblock \bibinfo{title}{Circuit {QED}: superconducting qubits coupled to
		microwave photons}.
	\newblock In \emph{\bibinfo{booktitle}{Quantum {Machines}: {Measurement} and
			{Control} of {Engineered} {Quantum} {Systems}}} (\bibinfo{publisher}{Oxford
		University Press}, \bibinfo{address}{Oxford}, \bibinfo{year}{2014}).
	\newblock
	\urlprefix\url{https://oxford.universitypressscholarship.com/view/10.1093/acprof:oso/9780199681181.001.0001/acprof-9780199681181-chapter-3}.
	
	\bibitem{nigg2012black}
	\bibinfo{author}{Nigg, S.~E.} \emph{et~al.}
	\newblock \bibinfo{title}{Black-{Box} {Superconducting} {Circuit}
		{Quantization}}.
	\newblock \emph{\bibinfo{journal}{Physical Review Letters}}
	\textbf{\bibinfo{volume}{108}}, \bibinfo{pages}{240502}
	(\bibinfo{year}{2012}).
	\newblock
	\urlprefix\url{https://link.aps.org/doi/10.1103/PhysRevLett.108.240502}.
	
	\bibitem{solgun2014blackbox}
	\bibinfo{author}{Solgun, F.}, \bibinfo{author}{Abraham, D.~W.} \&
	\bibinfo{author}{DiVincenzo, D.~P.}
	\newblock \bibinfo{title}{Blackbox quantization of superconducting circuits
		using exact impedance synthesis}.
	\newblock \emph{\bibinfo{journal}{Physical Review B}}
	\textbf{\bibinfo{volume}{90}}, \bibinfo{pages}{134504}
	(\bibinfo{year}{2014}).
	\newblock \urlprefix\url{https://link.aps.org/doi/10.1103/PhysRevB.90.134504}.
	
	\bibitem{minev_energy-participation_2021}
	\bibinfo{author}{Minev, Z.~K.} \emph{et~al.}
	\newblock \bibinfo{title}{Energy-participation quantization of {Josephson}
		circuits}.
	\newblock \emph{\bibinfo{journal}{npj Quantum Information}}
	\textbf{\bibinfo{volume}{7}}, \bibinfo{pages}{1--11} (\bibinfo{year}{2021}).
	\newblock \urlprefix\url{https://www.nature.com/articles/s41534-021-00461-8}.
	
	\bibitem{minev2021circuit}
	\bibinfo{author}{Minev, Z.~K.}, \bibinfo{author}{McConkey, T.~G.},
	\bibinfo{author}{Takita, M.}, \bibinfo{author}{Corcoles, A.~D.} \&
	\bibinfo{author}{Gambetta, J.~M.}
	\newblock \bibinfo{title}{Circuit quantum electrodynamics ({cQED}) with modular
		quasi-lumped models}.
	\newblock \emph{\bibinfo{journal}{arXiv:2103.10344 [cond-mat,
			physics:quant-ph]}}  (\bibinfo{year}{2021}).
	\newblock \urlprefix\url{http://arxiv.org/abs/2103.10344}.
	
	\bibitem{leib_many-body_2015}
	\bibinfo{author}{Leib, M.}
	\newblock \emph{\bibinfo{title}{Many-{Body} {Physics} with {Circuit} {Quantum}
			{Electrodynamics}}}.
	\newblock Ph.D. thesis, \bibinfo{school}{Technische Universität München}
	(\bibinfo{year}{2015}).
	\newblock
	\urlprefix\url{http://mediatum.ub.tum.de/1241486?id=1241486&change_language=en}.
	
	\bibitem{hagberg_exploring_2008}
	\bibinfo{author}{Hagberg, A.~A.}, \bibinfo{author}{Schult, D.~A.} \&
	\bibinfo{author}{Swart, P.~J.}
	\newblock \bibinfo{title}{Exploring {Network} {Structure}, {Dynamics}, and
		{Function} using {NetworkX}}.
	\newblock In \bibinfo{editor}{Varoquaux, G.}, \bibinfo{editor}{Vaught, T.} \&
	\bibinfo{editor}{Millman, J.} (eds.) \emph{\bibinfo{booktitle}{Proceedings of
			the 7th {Python} in {Science} {Conference}}}, \bibinfo{pages}{11 -- 15}
	(\bibinfo{address}{Pasadena, CA USA}, \bibinfo{year}{2008}).
	\newblock
	\urlprefix\url{http://conference.scipy.org/proceedings/SciPy2008/paper_2/}.
	
	\bibitem{meurer_sympy_2017}
	\bibinfo{author}{Meurer, A.} \emph{et~al.}
	\newblock \bibinfo{title}{{SymPy}: symbolic computing in {Python}}.
	\newblock \emph{\bibinfo{journal}{PeerJ Computer Science}}
	\textbf{\bibinfo{volume}{3}}, \bibinfo{pages}{e103} (\bibinfo{year}{2017}).
	\newblock \urlprefix\url{https://peerj.com/articles/cs-103}.
	
	\bibitem{langford_circuit_2013}
	\bibinfo{author}{Langford, N.~K.}
	\newblock \bibinfo{title}{Circuit {QED} - {Lecture} {Notes}}.
	\newblock \emph{\bibinfo{journal}{arXiv:1310.1897 [cond-mat,
			physics:quant-ph]}}  (\bibinfo{year}{2013}).
	\newblock \urlprefix\url{http://arxiv.org/abs/1310.1897}.
	\newblock \bibinfo{note}{ArXiv: 1310.1897}.
	
	\bibitem{virtanen_scipy_2020}
	\bibinfo{author}{Virtanen, P.} \emph{et~al.}
	\newblock \bibinfo{title}{{SciPy} 1.0: fundamental algorithms for scientific
		computing in {Python}}.
	\newblock \emph{\bibinfo{journal}{Nature Methods}}
	\textbf{\bibinfo{volume}{17}}, \bibinfo{pages}{261--272}
	(\bibinfo{year}{2020}).
	\newblock \urlprefix\url{https://www.nature.com/articles/s41592-019-0686-2}.
	
	\bibitem{catelani_relaxation_2011}
	\bibinfo{author}{Catelani, G.}, \bibinfo{author}{Schoelkopf, R.~J.},
	\bibinfo{author}{Devoret, M.~H.} \& \bibinfo{author}{Glazman, L.~I.}
	\newblock \bibinfo{title}{Relaxation and frequency shifts induced by
		quasiparticles in superconducting qubits}.
	\newblock \emph{\bibinfo{journal}{Physical Review B}}
	\textbf{\bibinfo{volume}{84}}, \bibinfo{pages}{064517}
	(\bibinfo{year}{2011}).
	\newblock \urlprefix\url{https://link.aps.org/doi/10.1103/PhysRevB.84.064517}.
	
	\bibitem{fernandes_lecture_nodate}
	\bibinfo{author}{Fernandes, R.~M.}
	\newblock \bibinfo{title}{Lecture {Notes}: {BCS} theory of superconductivity}.
	\newblock
	\urlprefix\url{https://portal.ifi.unicamp.br/images/files/graduacao/aulas-on-line/fen-emerg/lecture_notes_BCS.pdf}.
	
	\bibitem{cochran_superconducting_1958}
	\bibinfo{author}{Cochran, J.~F.} \& \bibinfo{author}{Mapother, D.~E.}
	\newblock \bibinfo{title}{Superconducting {Transition} in {Aluminum}}.
	\newblock \emph{\bibinfo{journal}{Physical Review}}
	\textbf{\bibinfo{volume}{111}}, \bibinfo{pages}{132--142}
	(\bibinfo{year}{1958}).
	\newblock \urlprefix\url{https://link.aps.org/doi/10.1103/PhysRev.111.132}.
	
	\bibitem{smith_superconducting_2020}
	\bibinfo{author}{Smith, W.~C.}, \bibinfo{author}{Kou, A.},
	\bibinfo{author}{Xiao, X.}, \bibinfo{author}{Vool, U.} \&
	\bibinfo{author}{Devoret, M.~H.}
	\newblock \bibinfo{title}{Superconducting circuit protected by
		two-{Cooper}-pair tunneling}.
	\newblock \emph{\bibinfo{journal}{npj Quantum Information}}
	\textbf{\bibinfo{volume}{6}}, \bibinfo{pages}{1--9} (\bibinfo{year}{2020}).
	\newblock \urlprefix\url{https://www.nature.com/articles/s41534-019-0231-2}.
	
	\bibitem{pop_coherent_2014}
	\bibinfo{author}{Pop, I.~M.} \emph{et~al.}
	\newblock \bibinfo{title}{Coherent suppression of electromagnetic dissipation
		due to superconducting quasiparticles}.
	\newblock \emph{\bibinfo{journal}{Nature}} \textbf{\bibinfo{volume}{508}},
	\bibinfo{pages}{369--372} (\bibinfo{year}{2014}).
	\newblock \urlprefix\url{http://www.nature.com/articles/nature13017}.
	
	\bibitem{yan_flux_2016}
	\bibinfo{author}{Yan, F.} \emph{et~al.}
	\newblock \bibinfo{title}{The flux qubit revisited to enhance coherence and
		reproducibility}.
	\newblock \emph{\bibinfo{journal}{Nature Communications}}
	\textbf{\bibinfo{volume}{7}}, \bibinfo{pages}{12964} (\bibinfo{year}{2016}).
	\newblock \urlprefix\url{https://www.nature.com/articles/ncomms12964}.
	
	\bibitem{lehoucq_arpack_1998}
	\bibinfo{author}{Lehoucq, R.~B.}, \bibinfo{author}{Sorensen, D.~C.} \&
	\bibinfo{author}{Yang, C.}
	\newblock \emph{\bibinfo{title}{{ARPACK} {Users}' {Guide}}}.
	\newblock Software, {Environments} and {Tools} (\bibinfo{publisher}{Society for
		Industrial and Applied Mathematics}, \bibinfo{year}{1998}).
	\newblock
	\urlprefix\url{https://epubs.siam.org/doi/book/10.1137/1.9780898719628}.
	
	\bibitem{martinis_energy_2009}
	\bibinfo{author}{Martinis, J.~M.}, \bibinfo{author}{Ansmann, M.} \&
	\bibinfo{author}{Aumentado, J.}
	\newblock \bibinfo{title}{Energy {Decay} in {Superconducting}
		{Josephson}-{Junction} {Qubits} from {Nonequilibrium} {Quasiparticle}
		{Excitations}}.
	\newblock \emph{\bibinfo{journal}{Physical Review Letters}}
	\textbf{\bibinfo{volume}{103}}, \bibinfo{pages}{097002}
	(\bibinfo{year}{2009}).
	\newblock
	\urlprefix\url{https://link.aps.org/doi/10.1103/PhysRevLett.103.097002}.
	
	\bibitem{serniak_nonequilibrium_2019}
	\bibinfo{author}{Serniak, K.}
	\newblock \emph{\bibinfo{title}{Nonequilibrium {Quasiparticles} in
			{Superconducting} {Qubits}}}.
	\newblock Ph.D. thesis, \bibinfo{school}{Yale University}
	(\bibinfo{year}{2019}).
	\newblock
	\urlprefix\url{https://cpb-us-w2.wpmucdn.com/campuspress.yale.edu/dist/2/3627/files/2020/10/kyle_thesis.pdf}.
	
	\bibitem{krantz_quantum_2019}
	\bibinfo{author}{Krantz, P.} \emph{et~al.}
	\newblock \bibinfo{title}{A quantum engineer's guide to superconducting
		qubits}.
	\newblock \emph{\bibinfo{journal}{Applied Physics Reviews}}
	\textbf{\bibinfo{volume}{6}}, \bibinfo{pages}{021318} (\bibinfo{year}{2019}).
	\newblock \urlprefix\url{https://aip.scitation.org/doi/10.1063/1.5089550}.
	
	\bibitem{koch_charge-insensitive_2007}
	\bibinfo{author}{Koch, J.} \emph{et~al.}
	\newblock \bibinfo{title}{Charge-insensitive qubit design derived from the
		{Cooper} pair box}.
	\newblock \emph{\bibinfo{journal}{Physical Review A}}
	\textbf{\bibinfo{volume}{76}}, \bibinfo{pages}{042319}
	(\bibinfo{year}{2007}).
	\newblock \urlprefix\url{https://link.aps.org/doi/10.1103/PhysRevA.76.042319}.
	
	\bibitem{manucharyan_fluxonium_2009}
	\bibinfo{author}{Manucharyan, V.~E.}, \bibinfo{author}{Koch, J.},
	\bibinfo{author}{Glazman, L.~I.} \& \bibinfo{author}{Devoret, M.~H.}
	\newblock \bibinfo{title}{Fluxonium: {Single} {Cooper}-{Pair} {Circuit} {Free}
		of {Charge} {Offsets}}.
	\newblock \emph{\bibinfo{journal}{Science}} \textbf{\bibinfo{volume}{326}},
	\bibinfo{pages}{113--116} (\bibinfo{year}{2009}).
	\newblock \urlprefix\url{https://science.sciencemag.org/content/326/5949/113}.
	
	\bibitem{orlando_superconducting_1999}
	\bibinfo{author}{Orlando, T.~P.} \emph{et~al.}
	\newblock \bibinfo{title}{Superconducting persistent-current qubit}.
	\newblock \emph{\bibinfo{journal}{Physical Review B}}
	\textbf{\bibinfo{volume}{60}}, \bibinfo{pages}{15398--15413}
	(\bibinfo{year}{1999}).
	\newblock \urlprefix\url{https://link.aps.org/doi/10.1103/PhysRevB.60.15398}.
	
	\bibitem{place_new_2021}
	\bibinfo{author}{Place, A. P.~M.} \emph{et~al.}
	\newblock \bibinfo{title}{New material platform for superconducting transmon
		qubits with coherence times exceeding 0.3 milliseconds}.
	\newblock \emph{\bibinfo{journal}{Nature Communications}}
	\textbf{\bibinfo{volume}{12}}, \bibinfo{pages}{1779} (\bibinfo{year}{2021}).
	\newblock \urlprefix\url{https://www.nature.com/articles/s41467-021-22030-5}.
	\newblock \bibinfo{note}{Number: 1 Publisher: Nature Publishing Group}.
	
	\bibitem{brooks_protected_2013}
	\bibinfo{author}{Brooks, P.}, \bibinfo{author}{Kitaev, A.} \&
	\bibinfo{author}{Preskill, J.}
	\newblock \bibinfo{title}{Protected gates for superconducting qubits}.
	\newblock \emph{\bibinfo{journal}{Physical Review A}}
	\textbf{\bibinfo{volume}{87}}, \bibinfo{pages}{052306}
	(\bibinfo{year}{2013}).
	\newblock \urlprefix\url{https://link.aps.org/doi/10.1103/PhysRevA.87.052306}.
	
	\bibitem{menke_automated_2021}
	\bibinfo{author}{Menke, T.} \emph{et~al.}
	\newblock \bibinfo{title}{Automated design of superconducting circuits and its
		application to 4-local couplers}.
	\newblock \emph{\bibinfo{journal}{npj Quantum Information}}
	\textbf{\bibinfo{volume}{7}}, \bibinfo{pages}{1--8} (\bibinfo{year}{2021}).
	\newblock \urlprefix\url{https://www.nature.com/articles/s41534-021-00382-6}.
	
	\bibitem{danilin_quantum_2021}
	\bibinfo{author}{Danilin, S.} \& \bibinfo{author}{Weides, M.}
	\newblock \bibinfo{title}{Quantum sensing with superconducting circuits}
	(\bibinfo{year}{2021}).
	\newblock \urlprefix\url{http://arxiv.org/abs/2103.11022}.
	\newblock \bibinfo{note}{Number: arXiv:2103.11022 arXiv:2103.11022 [quant-ph]}.
	
	\bibitem{kerremans_probabilistically_2022}
	\bibinfo{author}{Kerremans, T.}, \bibinfo{author}{Samuelsson, P.} \&
	\bibinfo{author}{Potts, P.}
	\newblock \bibinfo{title}{Probabilistically violating the first law of
		thermodynamics in a quantum heat engine}.
	\newblock \emph{\bibinfo{journal}{SciPost Physics}}
	\textbf{\bibinfo{volume}{12}}, \bibinfo{pages}{168} (\bibinfo{year}{2022}).
	\newblock \urlprefix\url{https://scipost.org/SciPostPhys.12.5.168}.
	
\end{thebibliography}
\end{document}